\documentclass[twocolumn,aps,prl,superscriptaddress,notitlepage,longbibliography]{revtex4}
\usepackage[colorlinks=true,urlcolor=blue,citecolor=blue,linkcolor=blue]{hyperref}
\usepackage{graphicx}  
\usepackage{dcolumn}   
\usepackage{bm}        
\usepackage{amssymb}
\usepackage{amsfonts}
\usepackage{doi}
\usepackage{verbatim}
\usepackage{epstopdf}
\usepackage{lineno}
\usepackage{amsmath}
\usepackage{braket}
\usepackage{hyperref}
\DeclareMathOperator{\Tr}{Tr}
\begin{document}
\title{Quantum Anomalous Layer Hall Effect in the Topological Magnet $\mathrm{MnBi_{2}Te_{4}}$}
\author{Wen-Bo Dai}
\affiliation{International Center for Quantum Materials, School of Physics, Peking University, Beijing 100871, China}
\affiliation{Beijing Academy of Quantum Information Sciences, Beijing 100193, China}
\author{Hailong Li}
\affiliation{International Center for Quantum Materials, School of Physics, Peking University, Beijing 100871, China}
\author{Dong-Hui Xu}
\affiliation{Department of Physics, Chongqing University, Chongqing 400044, China}
\author{Chui-Zhen Chen}
\email{czchen@suda.edu.cn}
\affiliation{School of Physical Science and Technology, Soochow University, Suzhou 215006, China}
\affiliation{Institute for Advanced Study, Soochow University, Suzhou 215006, China}
\author{X. C. Xie}
\email{xcxie@pku.edu.cn}
\affiliation{International Center for Quantum Materials, School of Physics, Peking University, Beijing 100871, China}
\affiliation{Collaborative Innovation Center of Quantum Matter, Beijing 100871, China}
\affiliation{CAS Center for Excellence in Topological Quantum Computation, University of Chinese Academy of Sciences, Beijing 100190, China}
\affiliation{Beijing Academy of Quantum Information Sciences, Beijing 100193, China}
\date{\today }

\begin{abstract}
Recently, a type of Hall effect due to an unusual layer-locked Berry curvature called the layer Hall effect (LHE) has been reported in the even-layered two-dimensional antiferromagnetic (AFM) $\mathrm{MnBi_{2}Te_{4}}$ [\textit{A. Gao et.al, Nature 595, 521 (2021)}].
In this work, we report that the quantization of LHE, which we call the quantum anomalous layer Hall effect (QALHE), can be realized in $\mathrm{MnBi_{2}Te_{4}}$. The QALHE originates from kicking a layer-locked Berry-curvature monopole out of the Fermi sea by a vertical electric field. Remarkably, we demonstrate that the electric-field reversal can switch the sign of the quantized Hall conductance of QALHE in the even-layered AFM phase. The QALHE can also be realized in the ferromagnetic phase. These results provide a promising way toward the electric engineering of the Berry curvature monopoles and quantized-layered transport in topological magnets.
\end{abstract}

\maketitle

{\emph{Introduction.}}---The quantum Hall effect observed in strong magnetic fields is one of most striking phenomena in condensed matter physics \cite{Klitzing1980,Thouless1982}. 
Recently, the notion of the quantum Hall effect is generalized to quantum spin Hall effect \cite{Kane2006A,Kane2006B,Bernevig2006,Konig2007,Du2015,Wu2018} and quantum valley Hall effect \cite{Xiao2007,Gorbachev2014} by utilizing spin and valley degrees of freedom.
They can be used to transport spin and valley current without dissipation, having potential applications in designing low-power devices. 
Yet, the electrons of opposite spin or valley indices in real materials spatially overlap with each other, leading to inevitable backscattering and thus short lifetimes of electron states \cite{Konig2007,Wu2018,Komatsu2018}, which hinders the development of these areas.
Therefore, a new class of robust quantum Hall effect characterized by spatial-resolved topological index is highly desirable.

\begin{figure}[bht]
\centering
\includegraphics[width=3.3in]{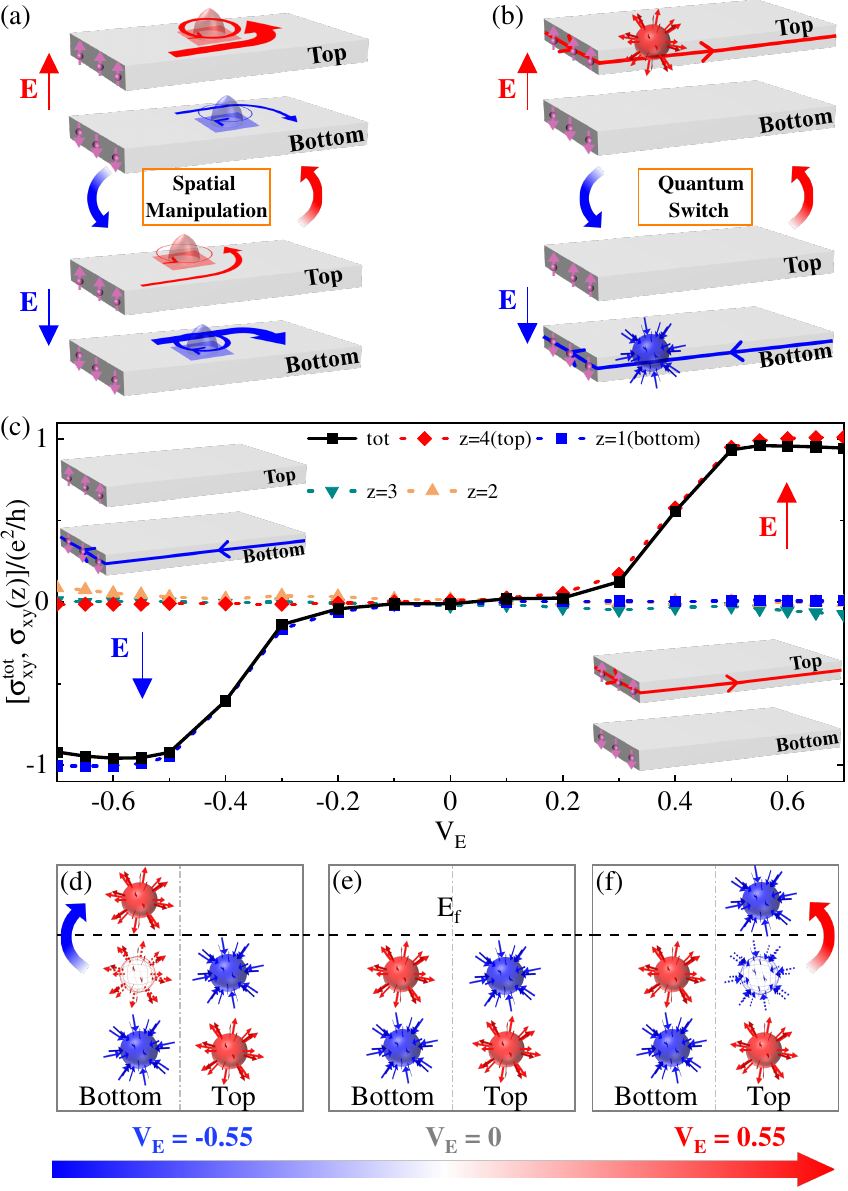}
\caption{(Color online). (a-b) Electric manipulation of the LHE and QALHE in even-layered AFM $\mathrm{MnBi_{2}Te_{4}}$.
(a) Opposite Berry curvature locked to the top and bottom layers, which are depleted in different layers under vertical $E$ field. Thus, the electrons deflecting in opposite
directions cannot cancel, creating a net anomalous Hall current \cite{LHE,supp}.
(b) Quantum switch of the QALHE (quantization  of the LHE) with opposite layer-locked Berry curvature monopoles, under vertical $E$ field.
(c) The Hall conductance of the whole system (solid line) $\sigma_{xy}^{tot}$ and of each layer (dash lines) $\sigma_{xy}(z)$ versus the vertical $E$-field strength $V_E$ in a four-layer AFM $\mathrm{MnBi_{2}Te_{4}}$. The Fermi energy is $E_f=1.2$ and system size $N_x=N_y=48$.
The parameters are set as $g^{-}_z=(-1)^z$, $M_{0}=-0.22, M_{1,2}=-0.22, A =0.78, A_{m}=0.44, A_{z}=0.1, B_0=1.1, B_{1,2}=0.92, B^{z}_{0}=0.2$ \cite{Liantwist,LiuQH} and other parameters equal to zero unless specified.
(d-f) Schematic plots of the layer-locked positive (red sphere) and negative (blue sphere) Berry curvature monopoles for different $V_E$.
The empty spheres in (d) and (f) indicate $E$-field-triggered Berry curvature monopoles transition across the Fermi level $E_f$.
\label{fig1} }
\end{figure}

Recently, significant progresses have been achieved to realize layered topologically magnetic systems \cite{Gong_2019,Rienks_2019,Otrokov_2019,Zhang_2019,Li_2019,Zhao_2020,Zhao_2021}.
In particular, the intrinsic antiferromagnetic (AFM) topological insulator (TI) $\mathrm{MnBi_{2}Te_{4}}$ has become a highly tunable platform to realize various of exotic topological phenomena due to the interplay between the Berry phase and its rich internal magnetic structures \cite{Deng_2020,GeWangJ,Wang.Y,Das,Liantwist,LiuQH,LHLlocallaxion}.
Notably, an intriguing Hall effect named the layer Hall effect (LHE) \cite{LHE} has been identified in the even-layered AFM $\mathrm{MnBi_{2}Te_{4}}$.
As shown in Fig.~\ref{fig1}(a), the LHE manifests emerging layer-dependent Hall current flowing in different directions,
because opposite Berry curvature is locked to top and bottom layers of $\mathrm{MnBi_{2}Te_{4}}$.
If the net Berry curvature is generated under an electric field, a layer-polarized anomalous Hall effect (AHE) arises in the LHE system.
This creates a new pathway to the spatial-tailoring of the Berry curvature via electric manipulation.
It is natural to conceive of further generalizing the LHE to its quantized version.

In this work, we propose layer-polarized quantum anomalous Hall effect (QAHE) in the disordered topological magnet $\mathrm{MnBi_{2}Te_{4}}$ under electric field, where a net Berry curvature monopole is locked to top or bottom layer [see Fig.~\ref{fig1}(b)].
This can be regarded as a quantized version of the LHE, so we call it the quantum anomalous layer Hall effect (QALHE).
Remarkably, in the even-layered AFM $\mathrm{MnBi_{2}Te_{4}}$, we demonstrate that the electric-field reversal can switch the sign of the quantized total Hall conductance,
where the quantized layer-dependent Hall conductance is switched between top and bottoms layers [see Fig.~\ref{fig1}(c)].
That's because that an electric field along the $z$-axis can transfer a net layer-locked topological monopole to above the Fermi level on the top or bottom layer [Figs.~\ref{fig1}(d) and (f)].
Here, the Berry-curvature monopole manifests as disorder-induced quantized Berry curvature in the energy space,
which are locked to the layer index [Fig.~\ref{fig1}(e)] and will shift in the energy space under electric fields.
To capture the underlying physics, we investigate the evolution of the Berry curvature and the ratio of the geometric mean density of states (DOS) $\rho_{typ}^{}$ to the arithmetic mean DOS $\rho_{avr}^{}$.
In addition, we show that QALHE can also be realized in the ferromagnetic (FM) $\mathrm{MnBi_{2}Te_{4}}$.
{\em Model Hamiltonian.}---The disordered $\mathrm{MnBi_{2}Te_{4}}$ under $E$ field can be described by a $4\times4$ effective Hamiltonian $H \!=\!H_{N}({\bf k})+g^{\mu}_zH_m({\bf k})+U(z)$ \cite{Liantwist,LiuQH,LHLlocallaxion}.
Here, the nonmagnetic part and the magnetic part $g^{\mu}_z H_m$ in momentum space are respectively given by
\begin{eqnarray}
 H_{N} \!&=&\!\mathcal{E} (\bf k)+\left(
    \begin{array}{cccc}
      m_{0}(\bf k) & A_z k_z & 0 & Ak_- \\
     A_{z}k_{z} & -m_{0}(\bf k) & Ak_-& 0 \\
      0 & Ak_+ & m_{0}(\bf k) & -A_zk_z \\
      Ak_+ & 0 & -A_z k_z & -m_{0}(\bf k) \\
   \end{array}
  \right)\nonumber
\end{eqnarray}
and
\begin{eqnarray*}
  H_{m}\!&=&\!\left(
    \begin{array}{cccc}
      m_{1}(\bf k) & 0 & 0 & A_m k_- \\
      0 & m_{2}(\bf k) & -A_m k_-& 0 \\
      0 & -A_m k_+ & -m_{1}(\bf k) & 0 \\
      A_mk_+ & 0 & 0 & -m_{2}(\bf k) \\
   \end{array}
  \right)\nonumber
\end{eqnarray*}
where $g^{+}_z=1$ and $g^{-}_z=(-1)^{z}$ describe the FM and AFM order, respectively.
Here, the wave vectors $k_{\pm}=k_x\pm ik_y$, $\mathcal{E} ({\bf k})=D_z k_z^2+D (k_x^2+k_y^2)$, $m_\nu ({\bf k})=M_{\nu} + B^{z}_{\nu} k_z^2 +B_\nu (k_x^2+k_y^2) $.
$U(z)=V_E[-(N_{z}+1)/2+z]$ represents the $z$th-layer potential in $N_z$ the layers sample caused by the electric-field along z-axis.
The magnetic disorder is included as $H_{W}({\bf r})=V({\bf r})\sigma_{z}$, 
where the random potential $V({\bf r})\in [-W/2,W/2]$ and the disorder strength $W=3.5$ unless specified.
The Pauli matrix $\sigma_{z}$ acts on the spin subspace.

{\em $E$-field triggered Berry curvature monopoles and QALHE.}---Due to the parity-time ($\mathcal{P} \mathcal{T} $) symmetry, the total Hall current of the even-layered AFM $\mathrm{MnBi_{2}Te_{4}}$ system vanishes without external fields \cite{supp}. However, unlike nonmagnetic topological systems,
an electric field can induce net Berry curvature and Hall currents in the AFM TI by breaking the $\mathcal{P} \mathcal{T} $ symmetry.
To investigate the $E$-field-induced Hall effect, we evaluate the total Hall conductance $\sigma_{xy}^{tot}=\sum_{z}\sigma_{xy}(z)$, 
where the layer-dependent Hall conductance in the $z$-th layer $\sigma_{xy}(z)=C_ze^2/h$ and the layer-dependent Chern number $C_z$ is given by \cite{Prodan_2009,Prodan_2011} 
\begin{eqnarray}
  \begin{array}{c}
    C_z=2\pi i \langle \Tr\{ P_{E_f} [-i[\hat{x},P_{E_f}],-i[\hat{y},P_{E_f}]]\}_z\rangle_{W},
  \end{array}
\end{eqnarray}
with $P_{E_f}$ being the projector onto the occupied states of $H$ below the Fermi energy $E_f$. $\hat{x}$ ($\hat{y}$) is coordinate operator, and
$\langle...\rangle_{W}$ means averaging over different disorder configurations.

Figure \ref{fig1}(c) plots the Hall conductance versus the $E$-field strength $V_E$.
Remarkably, it is found that the electric-field reversal can flip the Hall conductance $\sigma_{xy}^{tot}$ plateaus of the QAHE (see the black line).
This is closely analogous to the electric-field-reversible anomalous Hall effect
experimentally observed in the LHE system \cite{LHE,supp}, so we call it QALHE.
Further, we compare the total Hall conductance  $\sigma_{xy}^{tot}$ with the layer-dependent Hall conductance  $\sigma_{xy}(z)$.
For $V_E>0$, the $\sigma_{xy}^{tot}=e^2/h$ plateau is attributed to the $\sigma_{xy}(z=4)$ of the top layer (see the red line),
while (for $V_E<0$) the $\sigma_{xy}^{tot}=-e^2/h$ plateau comes from the $\sigma_{xy}(z=1$) of the bottom layer (see the blue line).
This means that the increase in upward (+$z$) and downward (-$z$) electric field can drive top and bottom layer from a topologically trivial phase into the QAHE phase of $\sigma_{xy}(z)=\pm e^2/h$, respectively.
Since the Hall conductance arises from the total Berry curvature of all the occupied states,
there exist Berry curvature monopoles under the Fermi level in the QALHE.
To this end, we provide a phenomenological explanation by using Berry curvature monopoles.
In Fig.~\ref{fig1}(e), at the zero electric field ($V_E=0$),
there exist a pair of positive and negative Berry curvature monopoles and the $\mathcal{P} \mathcal{T} $-symmetric partner
on the top and bottom layers, respectively.
Here two degenerate Berry curvature monopoles are of opposite values on the top and bottom layers due to the $\mathcal{P} \mathcal{T} $ symmetry.
As shown in Fig.\ref{fig1}(f), the monopoles of the top layer are lifted up in the energy space due to the upward electric field ($V_E>0$),
and they are transferred to above the Fermi energy $E_f$ when $V_E>0.55$ leaving a net ($+1$) Berry curvature monopole in the top layer and the whole system under $E_f$.
This explains why the upward electric field can drive the top-layer Hall conductance from $\sigma_{xy}(z)=0$ into $\sigma_{xy}(z)=1$ phase.
A similar process can give rise to a net ($-1$) Berry curvature monopole in the bottom layer and the whole system under a negative electric field [see Fig.\ref{fig1}(d)].

\begin{figure}[bht]
\centering
\includegraphics[width=3.3in]{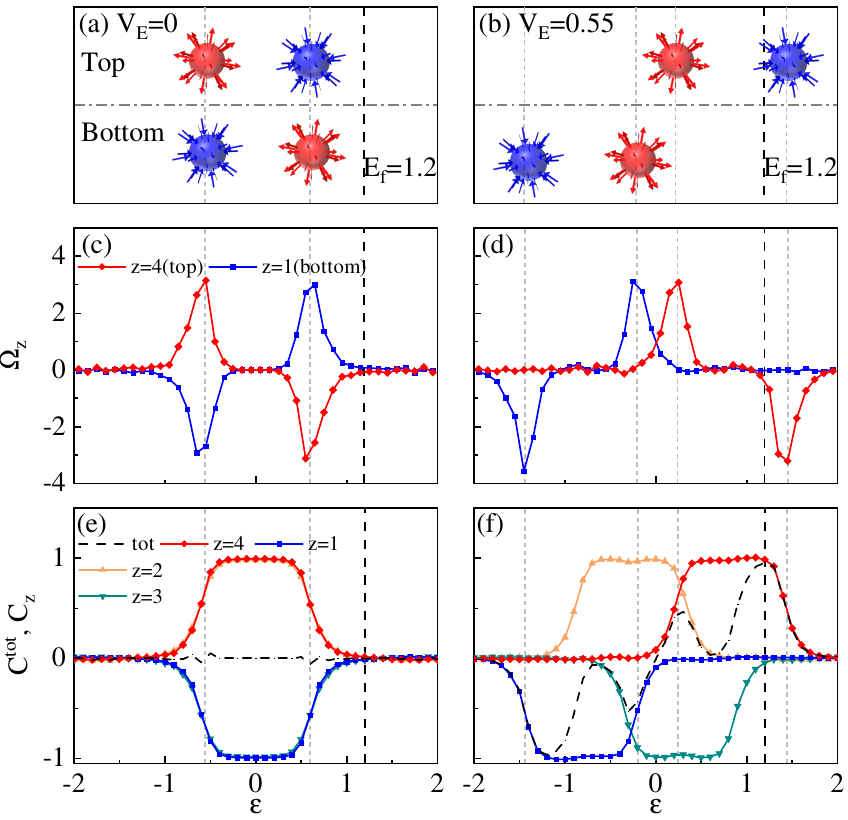}
\caption{(Color online).
Evolutions of Berry-curvature monopoles in AFM $\mathrm{MnBi_{2}Te_{4}}$ when the $E$ field increases from (a) $V_E =0$  to (b) $V_E =0.55$. Corresponding, (c-d) the layer-dependent Berry curvature $\Omega_z$ vs the energy $\varepsilon$,
and (e-f) the layer-dependent $C_z$ and the total Chern numbers $C^{tot}=\sum_z{C_z}$. The Fermi energy $E_f=1.2$ is fixed with the size $48\times 48 \times 4$.
\label{fig2} }
\end{figure}
To quantitatively elucidate the evolution of the Berry curvature monopoles under the electric field,
we calculated the layer-locked Berry curvature in energy space according to $C_z(\varepsilon)=\int_{-\infty}^{\varepsilon} \Omega _z(\epsilon ) \,d\mathcal{\epsilon }$ \cite{CCZ2021evolution},
where $\Omega_z(\varepsilon  )$ is the Berry curvature  of the z-th layer in energy space $\varepsilon$.
Figure \ref{fig2} plot the Berry curvature and Chern numbers, and sketch the corresponding Berry curvature monopole.
In Fig.\ref{fig2}(c) when $V_E=0$, the top-layer-locked Berry curvature $\Omega_{z=4(top)}$ has two peaks for each layer,
one positive and one negative, in the whole energy band.
Each peak corresponds to a topological phase transition between $C_z=\pm 1$ to $C_z=0$ at a critical energy $\varepsilon_c$ [see Fig.~\ref{fig2}(e)],
which will become a delta function $\delta(\varepsilon-\varepsilon_c)$ in the thermodynamic limit.
So each positive (negative) Berry curvature peak
manifests as a Berry curvature monopole with charge $+1$ ($-1$) in energy space as shown in Fig.~\ref{fig2}(a).

In Fig.~\ref{fig2}(a), there exists one pair of Berry curvature monopoles for the top and bottom layers each below the Fermi energy $E_f=1.2$, in accordance with those in Fig.~\ref{fig1}(e).
By increasing $V_E$ to $0.55$ in Fig.\ref{fig2} (b),
the top layer-locked Berry curvature monopoles (peaks) ascend to the $+\varepsilon $ direction,
while the bottom layer-locked monopoles (peaks) descend to the $-\varepsilon$ direction in the energy space,
due to the layer-dependent potential $U(z)=V_E[-(N_{z}+1)/2+z]$.
When a negative monopole in the top layer moves across the Fermi surface [see the blue ball in Fig.~\ref{fig2}(b)],
the AFM system reaches a top-layer-polarized QAHE phase with $C^{tot}=C_{z=4(top)}=1$ [see Fig.~\ref{fig2}(f)],  since there are only one positive monopole from the occupied states in the top layer
and no net monopoles in the other layers.
This explains the $E$-field-induced monopoles transition discussed in Fig.\ref{fig1}(f).
Similarly, an downward $E$ field will transfer a positive monopole in the bottom layer to above the Fermi energy [see Fig.\ref{fig1}(d)] due to the $\mathcal{P} \mathcal{T}$ symmetry \cite{supp}.

\begin{figure}[thb]
  \centering
  \includegraphics[width=3.3in]{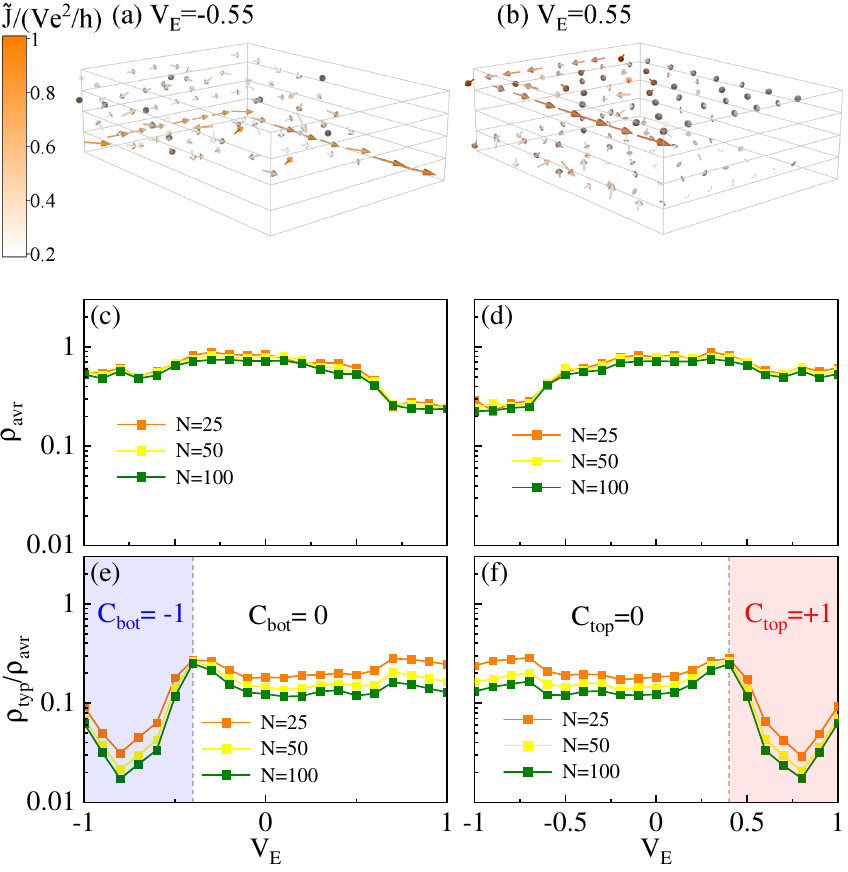}
  \caption{(Color online). Local current of the QALH when the $E$-field is reversed from (a) $V_E=-0.55$ to (b) $0.55$,
   with the system size $64\times 64 \times 4$.
  (c-d) The arithmetic mean DOS $\rho_{ave}$, and (e-f) the ratio of the geometric mean DOS $\rho_{typ}$ to $\rho_{ave}$ for the bottom layer and the top layer, respectively. The different colored lines indicate different system sizes $N\times N\times4$.
  Other parameters are the same as Fig.~\ref{fig2}.
  \label{fig3} }
  \end{figure}
{\em Quantum switch and scaling behavior.}---
The $E$-field switchable QALHE is accompanied by the switch of chiral edge channels on the boundary
and layer-resolved topological phase transitions in the bulk [ see Fig.~\ref{fig1}(b)].
To verify this point, we first evaluate the local current using the recursive Green's function method \cite{Jiang2009}, and
then calculate the geometric mean DOS $\rho_{typ}$ and the arithmetic mean DOS $\rho_{ave}$ during the phase transition process.
Here $\rho_{typ}$ and $\rho_{ave}$ are defined as \cite{ZYYgap,ZYYDOSTAI,Dobrosavljevi__2003,Schubert,JANSSEN19981}
\begin{eqnarray}
  \begin{array}{c}
    \rho_{ave}(E_f)=\langle \langle  \rho(i,E_f)\rangle \rangle \\
    \\
    \rho_{typ}(E_f)=\exp[\langle \langle \ln\rho(i,E_f)\rangle \rangle]
  \end{array}
\end{eqnarray}
where $\langle \langle \cdots \rangle \rangle $ denotes the arithmetic average over the sample sites and disorder realizations.
The local DOS on the Fermi energy $E_f$ is determined by
$\rho(i,E_f)=\sum_{l,n,\alpha}|\left\langle il | n\alpha\right\rangle |^2\delta(E_f-E_n)$
where $i$, $l$ denotes the site index, orbital index and $n$ is the eigenvalue index.
To calculate $\rho_{typ}$ and $\rho_{ave}$, we approximate $\delta(x)\approx \eta /[\pi(x^2 + \eta ^2)]$ with $\eta = 10^{-4}$,
and diagonalize the lattice Hamiltonian with periodic boundary conditions in $x$ and $y$ directions.
For an extended state that uniformly distributes over the sample, $\rho_{typ}$ is almost the same as $\rho_{ave}$.
In contrast, for a localized state concentrated on certain sites, $\rho_{typ}$ will be extremely small.
The ratio $\rho_{typ}/\rho_{ave}$ keeps finite for extended states, while $\rho_{typ}/\rho_{ave}$ approaches zero for localized states, in the thermodynamic limit.

Figures \ref{fig3}~(a)-(b) show the local current of the QALHE at $E_f=1.2$. When the $E$ field is reversed from $V_E=-0.55$ to $0.55$,
the chiral edge mode is switched from the bottom to top layer.
The nonzero local current in the bulk indicates the existence of the disorder-induced localized states.
Also, the finite scaling-independent $\rho_{ave}$ in Figs.~\ref{fig3}(c-d) and vanishing  $\rho_{typ}/\rho_{ave}$ with increasing the size $N$ in Figs.~\ref{fig3}(e-f) at $V_E=\pm0.5$ verify the existence of localized states.
Furthermore,  in Fig.~\ref{fig3}(f), there exists one fixed point for the top layer at $V_E^c\approx 0.4$ where $\rho_{typ}/\rho_{ave}$ is independent of $N$.
This is coincident with layer-resolved transition point from $C_{top}=0$ to $C_{top}=1$ of the top layer [see the red line] in Fig.~\ref{fig1}(c),
where the Berry curvature monopole is crossing the Fermi level Fig.\ref{fig1}(f).
Similarly, the fixed point of $\rho_{typ}/\rho_{ave}$ for the bottom layer at $V_E^c\approx -0.4$ in Fig.~\ref{fig3}(e) agrees with the transition of the bottom layer [see the blue line] in Fig.~\ref{fig1}(c).
Therefore, the quantum switch process of the QALHE is accompanied by switching the layer-polarized edge states on the boundary and  the layer-dependent Anderson transitions in the bulk.

\begin{figure}[tbh]
  \centering
  \includegraphics[width=3.3in]{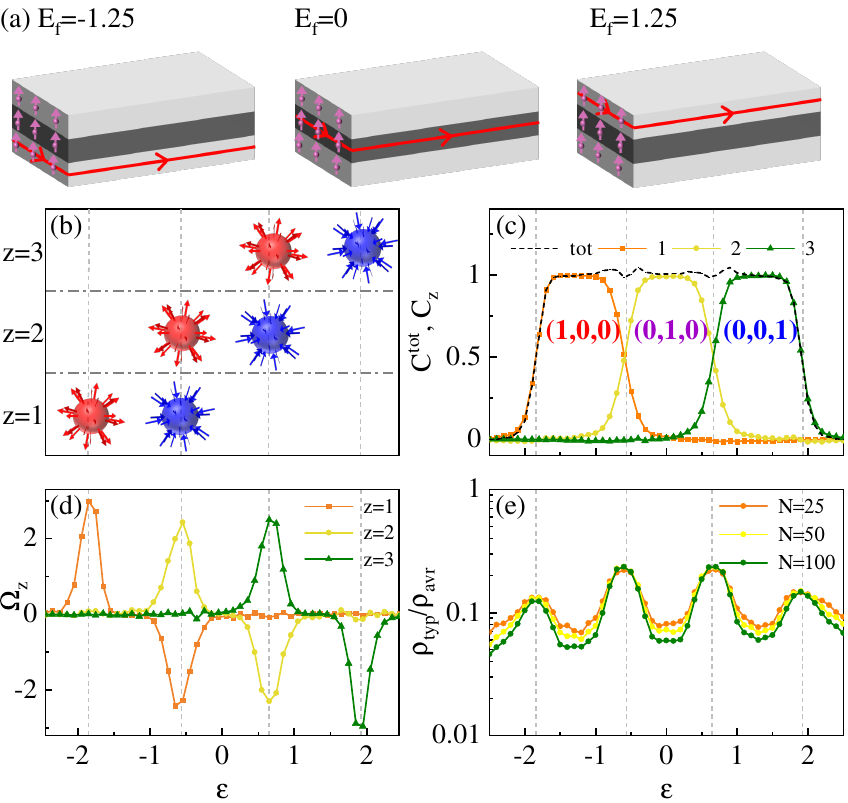}
  \caption{(Color online).  (a) Schematic plots of the QALHE in the FM $\mathrm{MnBi_{2}Te_{4}}$ with the Fermi energies $E_f=-1.25$, $0$  and $1.25$.
  (b) The layer-dependent Berry curvature monopoles,
  (c)  The layer-dependent and total Chern numbers, and (d) the layer-dependent Berry curvature VS the energy $\varepsilon$.
  (e) The ratio of the  geometric mean DOS to the  arithmetic mean DOS $\rho_{typ}/\rho_{ave}$ of the whole system for different system sizes $N\times N$.
  The dashed lines indicate that  the locations of critical points in (e), and Berry curvature monopoles in (b) and (d) coincide.
   We set  $g^{+}_z=1, M_{0}=-0.3, M_{1,2}=-0.1, A=0.78, A_{m}=0.46, A_{z}=0.1, B_0=1.1, B_{1,2}=0.9, B^{z}_{0}=0.2, D_z =0.06$ \cite{Liantwist}. The $E$ field strength $V_E=1.25$.
  \label{fig4} }
  \end{figure}
{\em QALHE and quantum switch in FM phase.}---Recently, the high Chern number QAHE was experimentally discovered in the FM MnBi$_2$Te$_4$ \cite{Ge_2020} and in the 2D-layered TI heterostructures \cite{Zhao_2020,Zhao_2021}, which suggests a new platform to realize richer topological phenomena.
As shown by Fig.~\ref{fig4}(c), the layer-dependent Chern numbers  $ (C_1,  C_2,  C_3)$
are tuned from $(1,  0,  0)$ to $(0,  1,  0)$ and then to $(0,  0,  1)$ by varying the energy  $\varepsilon=E_f=-1.25$, to $0$ and then to $1.25$, where the edge channel is switched from the bottom to the middle  and then to the top layer [see Figs.~\ref{fig4}(a)-(c)].
That's because the layer-dependent Chern numbers changes values when the Fermi energy goes across the discrete locations of the Berry curvature monopoles  Fig.~\ref{fig4}(b), which are determined by the layer-dependent Berry curvature in Fig.~\ref{fig4}(d).
Furthermore, we plot $\rho_{typ}/\rho_{ave}$ of the whole system versus energy in Fig.~\ref{fig4}(e),
and find that the locations of critical points and the Berry-curvature monopoles coincide.
These results suggest that the Anderson localization plays an important role in the quantum switch process of the QALHE.

In reality, the QAHLE can be realized in the much thicker FM topological magnet samples with higher Chern number (see the supplementary materials \cite{supp}) such as the FM MnBi$_2$Te$_4$ \cite{Ge_2020} and 2D-layered TI heterostructures \cite{Zhao_2020}.
Meanwhile, we propose that the quantum switch of the layer-dependent edge channel can be detected by a four-terminal device in experiments \cite{supp}.

{\em Discussions and conclusions.}---
The QAHLE can be viewed as a localization-driven counterpart to the recently reported LHE by A. Gao et.al \cite{LHE}.
As a comparison, we have shown the LHE in the clean or weak disordered AFM MnBi$_2$Te$_4$  \cite{supp}.
Moreover, the 3D AFM MnBi$_2$Te$_4$ is originally peridicted to be an axion insulator exhibiting topological magnetoelectricity effect when the Fermi energy is inside the band gap \cite{Zhang_2019,Li_2019}.
In contrast, to realize the QAHLE in the quasi-2D AFM MnBi$_2$Te$_4$, the Fermi energy is tuned to the 2D Anderson-localized bulk states.
In summary, we have revealed that the QALHE, a quantized version of the LHE, can be realized
in both the AFM and FM $\mathrm{MnBi_{2}Te_{4}}$. The QALHE is attributed to electric-tunable the Berry-curvature monopoles.
Further, we show that the electric field can switch the QALHE edge channel among different layers.
Our work paves the way to eletrically tunable layer-resolved disspationless transport in topological magnets.

{\emph{Acknowledgement.}}---  We thank Qing-Feng Sun for illuminating discussions.
This work was financially supported by National Key R and D Program of China (Grant No. 2017YFA0303301), NBRPC (Grant No. 2015CB921102), NSFC (Grants Nos. 11534001, 11822407, 11921005, 12074108), and also supported
by the Fundamental Research Funds for the Central Universities, the Strategic Priority Research Program of Chinese Academy of Sciences (DB28000000), and Beijing Municipal Science \& Technology
Commission (Grant No. Z191100007219013).
C.-Z.C. is also funded by the Priority Academic Program Development of Jiangsu Higher Education Institutions.


\begin{thebibliography}{37}%
  \makeatletter
  \providecommand \@ifxundefined [1]{%
   \@ifx{#1\undefined}
  }%
  \providecommand \@ifnum [1]{%
   \ifnum #1\expandafter \@firstoftwo
   \else \expandafter \@secondoftwo
   \fi
  }%
  \providecommand \@ifx [1]{%
   \ifx #1\expandafter \@firstoftwo
   \else \expandafter \@secondoftwo
   \fi
  }%
  \providecommand \natexlab [1]{#1}%
  \providecommand \enquote  [1]{``#1''}%
  \providecommand \bibnamefont  [1]{#1}%
  \providecommand \bibfnamefont [1]{#1}%
  \providecommand \citenamefont [1]{#1}%
  \providecommand \href@noop [0]{\@secondoftwo}%
  \providecommand \href [0]{\begingroup \@sanitize@url \@href}%
  \providecommand \@href[1]{\@@startlink{#1}\@@href}%
  \providecommand \@@href[1]{\endgroup#1\@@endlink}%
  \providecommand \@sanitize@url [0]{\catcode `\\12\catcode `\$12\catcode
    `\&12\catcode `\#12\catcode `\^12\catcode `\_12\catcode `\%12\relax}%
  \providecommand \@@startlink[1]{}%
  \providecommand \@@endlink[0]{}%
  \providecommand \url  [0]{\begingroup\@sanitize@url \@url }%
  \providecommand \@url [1]{\endgroup\@href {#1}{\urlprefix }}%
  \providecommand \urlprefix  [0]{URL }%
  \providecommand \Eprint [0]{\href }%
  \providecommand \doibase [0]{http://dx.doi.org/}%
  \providecommand \selectlanguage [0]{\@gobble}%
  \providecommand \bibinfo  [0]{\@secondoftwo}%
  \providecommand \bibfield  [0]{\@secondoftwo}%
  \providecommand \translation [1]{[#1]}%
  \providecommand \BibitemOpen [0]{}%
  \providecommand \bibitemStop [0]{}%
  \providecommand \bibitemNoStop [0]{.\EOS\space}%
  \providecommand \EOS [0]{\spacefactor3000\relax}%
  \providecommand \BibitemShut  [1]{\csname bibitem#1\endcsname}%
  \let\auto@bib@innerbib\@empty
  \bibitem [{\citenamefont {Klitzing}\ \emph {et~al.}(1980)\citenamefont
    {Klitzing}, \citenamefont {Dorda},\ and\ \citenamefont
    {Pepper}}]{Klitzing1980}%
    \BibitemOpen
    \bibfield  {author} {\bibinfo {author} {\bibfnamefont {K.~v.}\ \bibnamefont
    {Klitzing}}, \bibinfo {author} {\bibfnamefont {G.}~\bibnamefont {Dorda}}, \
    and\ \bibinfo {author} {\bibfnamefont {M.}~\bibnamefont {Pepper}},\ }\href
    {\doibase 10.1103/PhysRevLett.45.494} {\bibfield  {journal} {\bibinfo
    {journal} {Phys. Rev. Lett.}\ }\textbf {\bibinfo {volume} {45}},\ \bibinfo
    {pages} {494} (\bibinfo {year} {1980})}\BibitemShut {NoStop}%
  \bibitem [{\citenamefont {Thouless}\ \emph {et~al.}(1982)\citenamefont
    {Thouless}, \citenamefont {Kohmoto}, \citenamefont {Nightingale},\ and\
    \citenamefont {Den~Nijs}}]{Thouless1982}%
    \BibitemOpen
    \bibfield  {author} {\bibinfo {author} {\bibfnamefont {D.~J.}\ \bibnamefont
    {Thouless}}, \bibinfo {author} {\bibfnamefont {M.}~\bibnamefont {Kohmoto}},
    \bibinfo {author} {\bibfnamefont {M.~P.}\ \bibnamefont {Nightingale}}, \ and\
    \bibinfo {author} {\bibfnamefont {M.}~\bibnamefont {Den~Nijs}},\ }\href
    {\doibase 10.1103/PhysRevLett.49.405} {\bibfield  {journal} {\bibinfo
    {journal} {Phys. Rev. Lett.}\ }\textbf {\bibinfo {volume} {49}},\ \bibinfo
    {pages} {405} (\bibinfo {year} {1982})}\BibitemShut {NoStop}%
  \bibitem [{\citenamefont {Kane}\ and\ \citenamefont
    {Mele}(2005{\natexlab{a}})}]{Kane2006A}%
    \BibitemOpen
    \bibfield  {author} {\bibinfo {author} {\bibfnamefont {C.~L.}\ \bibnamefont
    {Kane}}\ and\ \bibinfo {author} {\bibfnamefont {E.~J.}\ \bibnamefont
    {Mele}},\ }\href {\doibase 10.1103/PhysRevLett.95.226801} {\bibfield
    {journal} {\bibinfo  {journal} {Phys. Rev. Lett.}\ }\textbf {\bibinfo
    {volume} {95}},\ \bibinfo {pages} {226801} (\bibinfo {year}
    {2005}{\natexlab{a}})}\BibitemShut {NoStop}%
  \bibitem [{\citenamefont {Kane}\ and\ \citenamefont
    {Mele}(2005{\natexlab{b}})}]{Kane2006B}%
    \BibitemOpen
    \bibfield  {author} {\bibinfo {author} {\bibfnamefont {C.~L.}\ \bibnamefont
    {Kane}}\ and\ \bibinfo {author} {\bibfnamefont {E.~J.}\ \bibnamefont
    {Mele}},\ }\href {\doibase 10.1103/PhysRevLett.95.146802} {\bibfield
    {journal} {\bibinfo  {journal} {Phys. Rev. Lett.}\ }\textbf {\bibinfo
    {volume} {95}},\ \bibinfo {pages} {146802} (\bibinfo {year}
    {2005}{\natexlab{b}})}\BibitemShut {NoStop}%
  \bibitem [{\citenamefont {Bernevig}\ \emph {et~al.}(2006)\citenamefont
    {Bernevig}, \citenamefont {Hughes},\ and\ \citenamefont
    {Zhang}}]{Bernevig2006}%
    \BibitemOpen
    \bibfield  {author} {\bibinfo {author} {\bibfnamefont {B.~A.}\ \bibnamefont
    {Bernevig}}, \bibinfo {author} {\bibfnamefont {T.~L.}\ \bibnamefont
    {Hughes}}, \ and\ \bibinfo {author} {\bibfnamefont {S.-C.}\ \bibnamefont
    {Zhang}},\ }\href {\doibase 10.1126/science.1133734} {\bibfield  {journal}
    {\bibinfo  {journal} {Science}\ }\textbf {\bibinfo {volume} {314}},\ \bibinfo
    {pages} {1757} (\bibinfo {year} {2006})}\BibitemShut {NoStop}%
  \bibitem [{\citenamefont {Konig}\ \emph {et~al.}(2007)\citenamefont {Konig},
    \citenamefont {Wiedmann}, \citenamefont {Brune}, \citenamefont {Roth},
    \citenamefont {Buhmann}, \citenamefont {Molenkamp}, \citenamefont {Qi},\ and\
    \citenamefont {Zhang}}]{Konig2007}%
    \BibitemOpen
    \bibfield  {author} {\bibinfo {author} {\bibfnamefont {M.}~\bibnamefont
    {Konig}}, \bibinfo {author} {\bibfnamefont {S.}~\bibnamefont {Wiedmann}},
    \bibinfo {author} {\bibfnamefont {C.}~\bibnamefont {Brune}}, \bibinfo
    {author} {\bibfnamefont {A.}~\bibnamefont {Roth}}, \bibinfo {author}
    {\bibfnamefont {H.}~\bibnamefont {Buhmann}}, \bibinfo {author} {\bibfnamefont
    {L.~W.}\ \bibnamefont {Molenkamp}}, \bibinfo {author} {\bibfnamefont {X.-L.}\
    \bibnamefont {Qi}}, \ and\ \bibinfo {author} {\bibfnamefont {S.-C.}\
    \bibnamefont {Zhang}},\ }\href@noop {} {\bibfield  {journal} {\bibinfo
    {journal} {Science}\ }\textbf {\bibinfo {volume} {318}},\ \bibinfo {pages}
    {766} (\bibinfo {year} {2007})}\BibitemShut {NoStop}%
  \bibitem [{\citenamefont {Du}\ \emph {et~al.}(2015)\citenamefont {Du},
    \citenamefont {Knez}, \citenamefont {Sullivan},\ and\ \citenamefont
    {Du}}]{Du2015}%
    \BibitemOpen
    \bibfield  {author} {\bibinfo {author} {\bibfnamefont {L.}~\bibnamefont
    {Du}}, \bibinfo {author} {\bibfnamefont {I.}~\bibnamefont {Knez}}, \bibinfo
    {author} {\bibfnamefont {G.}~\bibnamefont {Sullivan}}, \ and\ \bibinfo
    {author} {\bibfnamefont {R.-R.}\ \bibnamefont {Du}},\ }\href {\doibase
    10.1103/PhysRevLett.114.096802} {\bibfield  {journal} {\bibinfo  {journal}
    {Phys. Rev. Lett.}\ }\textbf {\bibinfo {volume} {114}},\ \bibinfo {pages}
    {096802} (\bibinfo {year} {2015})}\BibitemShut {NoStop}%
  \bibitem [{\citenamefont {Wu}\ \emph {et~al.}(2018)\citenamefont {Wu},
    \citenamefont {Fatemi}, \citenamefont {Gibson}, \citenamefont {Watanabe},
    \citenamefont {Taniguchi}, \citenamefont {Cava},\ and\ \citenamefont
    {Jarillo-Herrero}}]{Wu2018}%
    \BibitemOpen
    \bibfield  {author} {\bibinfo {author} {\bibfnamefont {S.}~\bibnamefont
    {Wu}}, \bibinfo {author} {\bibfnamefont {V.}~\bibnamefont {Fatemi}}, \bibinfo
    {author} {\bibfnamefont {Q.~D.}\ \bibnamefont {Gibson}}, \bibinfo {author}
    {\bibfnamefont {K.}~\bibnamefont {Watanabe}}, \bibinfo {author}
    {\bibfnamefont {T.}~\bibnamefont {Taniguchi}}, \bibinfo {author}
    {\bibfnamefont {R.~J.}\ \bibnamefont {Cava}}, \ and\ \bibinfo {author}
    {\bibfnamefont {P.}~\bibnamefont {Jarillo-Herrero}},\ }\href {\doibase
    10.1126/science.aan6003} {\bibfield  {journal} {\bibinfo  {journal}
    {Science}\ }\textbf {\bibinfo {volume} {359}},\ \bibinfo {pages} {76}
    (\bibinfo {year} {2018})}\BibitemShut {NoStop}%
  \bibitem [{\citenamefont {Xiao}\ \emph {et~al.}(2007)\citenamefont {Xiao},
    \citenamefont {Yao},\ and\ \citenamefont {Niu}}]{Xiao2007}%
    \BibitemOpen
    \bibfield  {author} {\bibinfo {author} {\bibfnamefont {D.}~\bibnamefont
    {Xiao}}, \bibinfo {author} {\bibfnamefont {W.}~\bibnamefont {Yao}}, \ and\
    \bibinfo {author} {\bibfnamefont {Q.}~\bibnamefont {Niu}},\ }\href {\doibase
    10.1103/PhysRevLett.99.236809} {\bibfield  {journal} {\bibinfo  {journal}
    {Phys. Rev. Lett.}\ }\textbf {\bibinfo {volume} {99}},\ \bibinfo {pages}
    {236809} (\bibinfo {year} {2007})}\BibitemShut {NoStop}%
  \bibitem [{\citenamefont {Gorbachev}\ \emph {et~al.}(2014)\citenamefont
    {Gorbachev}, \citenamefont {Song}, \citenamefont {Yu}, \citenamefont
    {Kretinin}, \citenamefont {Withers}, \citenamefont {Cao}, \citenamefont
    {Mishchenko}, \citenamefont {Grigorieva}, \citenamefont {Novoselov},
    \citenamefont {Levitov},\ and\ \citenamefont {Geim}}]{Gorbachev2014}%
    \BibitemOpen
    \bibfield  {author} {\bibinfo {author} {\bibfnamefont {R.~V.}\ \bibnamefont
    {Gorbachev}}, \bibinfo {author} {\bibfnamefont {J.~C.~W.}\ \bibnamefont
    {Song}}, \bibinfo {author} {\bibfnamefont {G.~L.}\ \bibnamefont {Yu}},
    \bibinfo {author} {\bibfnamefont {A.~V.}\ \bibnamefont {Kretinin}}, \bibinfo
    {author} {\bibfnamefont {F.}~\bibnamefont {Withers}}, \bibinfo {author}
    {\bibfnamefont {Y.}~\bibnamefont {Cao}}, \bibinfo {author} {\bibfnamefont
    {A.}~\bibnamefont {Mishchenko}}, \bibinfo {author} {\bibfnamefont {I.~V.}\
    \bibnamefont {Grigorieva}}, \bibinfo {author} {\bibfnamefont {K.~S.}\
    \bibnamefont {Novoselov}}, \bibinfo {author} {\bibfnamefont {L.~S.}\
    \bibnamefont {Levitov}}, \ and\ \bibinfo {author} {\bibfnamefont {A.~K.}\
    \bibnamefont {Geim}},\ }\href {\doibase 10.1126/science.1254966} {\bibfield
    {journal} {\bibinfo  {journal} {Science}\ }\textbf {\bibinfo {volume}
    {346}},\ \bibinfo {pages} {448} (\bibinfo {year} {2014})}\BibitemShut
    {NoStop}%
  \bibitem [{\citenamefont {Komatsu}\ \emph {et~al.}(2018)\citenamefont
    {Komatsu}, \citenamefont {Morita}, \citenamefont {Watanabe}, \citenamefont
    {Tsuya}, \citenamefont {Watanabe}, \citenamefont {Taniguchi},\ and\
    \citenamefont {Moriyama}}]{Komatsu2018}%
    \BibitemOpen
    \bibfield  {author} {\bibinfo {author} {\bibfnamefont {K.}~\bibnamefont
    {Komatsu}}, \bibinfo {author} {\bibfnamefont {Y.}~\bibnamefont {Morita}},
    \bibinfo {author} {\bibfnamefont {E.}~\bibnamefont {Watanabe}}, \bibinfo
    {author} {\bibfnamefont {D.}~\bibnamefont {Tsuya}}, \bibinfo {author}
    {\bibfnamefont {K.}~\bibnamefont {Watanabe}}, \bibinfo {author}
    {\bibfnamefont {T.}~\bibnamefont {Taniguchi}}, \ and\ \bibinfo {author}
    {\bibfnamefont {S.}~\bibnamefont {Moriyama}},\ }\href {\doibase
    10.1126/sciadv.aaq0194} {\bibfield  {journal} {\bibinfo  {journal} {Science
    Advances}\ }\textbf {\bibinfo {volume} {4}} (\bibinfo {year} {2018}),\
    10.1126/sciadv.aaq0194}\BibitemShut {NoStop}%
  \bibitem [{\citenamefont {Gong}\ \emph {et~al.}(2019)\citenamefont {Gong},
    \citenamefont {Guo}, \citenamefont {Li}, \citenamefont {Zhu}, \citenamefont
    {Liao}, \citenamefont {Liu}, \citenamefont {Zhang}, \citenamefont {Gu},
    \citenamefont {Tang}, \citenamefont {Feng}, \citenamefont {Zhang},
    \citenamefont {Li}, \citenamefont {Song}, \citenamefont {Wang}, \citenamefont
    {Yu}, \citenamefont {Chen}, \citenamefont {Wang}, \citenamefont {Yao},
    \citenamefont {Duan}, \citenamefont {Xu}, \citenamefont {Zhang},
    \citenamefont {Ma}, \citenamefont {Xue},\ and\ \citenamefont
    {He}}]{Gong_2019}%
    \BibitemOpen
    \bibfield  {author} {\bibinfo {author} {\bibfnamefont {Y.}~\bibnamefont
    {Gong}}, \bibinfo {author} {\bibfnamefont {J.}~\bibnamefont {Guo}}, \bibinfo
    {author} {\bibfnamefont {J.}~\bibnamefont {Li}}, \bibinfo {author}
    {\bibfnamefont {K.}~\bibnamefont {Zhu}}, \bibinfo {author} {\bibfnamefont
    {M.}~\bibnamefont {Liao}}, \bibinfo {author} {\bibfnamefont {X.}~\bibnamefont
    {Liu}}, \bibinfo {author} {\bibfnamefont {Q.}~\bibnamefont {Zhang}}, \bibinfo
    {author} {\bibfnamefont {L.}~\bibnamefont {Gu}}, \bibinfo {author}
    {\bibfnamefont {L.}~\bibnamefont {Tang}}, \bibinfo {author} {\bibfnamefont
    {X.}~\bibnamefont {Feng}}, \bibinfo {author} {\bibfnamefont {D.}~\bibnamefont
    {Zhang}}, \bibinfo {author} {\bibfnamefont {W.}~\bibnamefont {Li}}, \bibinfo
    {author} {\bibfnamefont {C.}~\bibnamefont {Song}}, \bibinfo {author}
    {\bibfnamefont {L.}~\bibnamefont {Wang}}, \bibinfo {author} {\bibfnamefont
    {P.}~\bibnamefont {Yu}}, \bibinfo {author} {\bibfnamefont {X.}~\bibnamefont
    {Chen}}, \bibinfo {author} {\bibfnamefont {Y.}~\bibnamefont {Wang}}, \bibinfo
    {author} {\bibfnamefont {H.}~\bibnamefont {Yao}}, \bibinfo {author}
    {\bibfnamefont {W.}~\bibnamefont {Duan}}, \bibinfo {author} {\bibfnamefont
    {Y.}~\bibnamefont {Xu}}, \bibinfo {author} {\bibfnamefont {S.-C.}\
    \bibnamefont {Zhang}}, \bibinfo {author} {\bibfnamefont {X.}~\bibnamefont
    {Ma}}, \bibinfo {author} {\bibfnamefont {Q.-K.}\ \bibnamefont {Xue}}, \ and\
    \bibinfo {author} {\bibfnamefont {K.}~\bibnamefont {He}},\ }\href {\doibase
    10.1088/0256-307x/36/7/076801} {\bibfield  {journal} {\bibinfo  {journal}
    {Chinese Physics Letters}\ }\textbf {\bibinfo {volume} {36}},\ \bibinfo
    {pages} {076801} (\bibinfo {year} {2019})}\BibitemShut {NoStop}%
  \bibitem [{\citenamefont {Rienks}\ \emph {et~al.}(2019)\citenamefont {Rienks},
    \citenamefont {Wimmer}, \citenamefont {S{\'{a}}nchez-Barriga}, \citenamefont
    {Caha}, \citenamefont {Mandal}, \citenamefont {R{\r{u}}{\v{z}}i{\v{c}}ka},
    \citenamefont {Ney}, \citenamefont {Steiner}, \citenamefont {Volobuev},
    \citenamefont {Groiss}, \citenamefont {Albu}, \citenamefont {Kothleitner},
    \citenamefont {Michali{\v{c}}ka}, \citenamefont {Khan}, \citenamefont
    {Min{\'{a}}r}, \citenamefont {Ebert}, \citenamefont {Bauer}, \citenamefont
    {Freyse}, \citenamefont {Varykhalov}, \citenamefont {Rader},\ and\
    \citenamefont {Springholz}}]{Rienks_2019}%
    \BibitemOpen
    \bibfield  {author} {\bibinfo {author} {\bibfnamefont {E.~D.~L.}\
    \bibnamefont {Rienks}}, \bibinfo {author} {\bibfnamefont {S.}~\bibnamefont
    {Wimmer}}, \bibinfo {author} {\bibfnamefont {J.}~\bibnamefont
    {S{\'{a}}nchez-Barriga}}, \bibinfo {author} {\bibfnamefont {O.}~\bibnamefont
    {Caha}}, \bibinfo {author} {\bibfnamefont {P.~S.}\ \bibnamefont {Mandal}},
    \bibinfo {author} {\bibfnamefont {J.}~\bibnamefont
    {R{\r{u}}{\v{z}}i{\v{c}}ka}}, \bibinfo {author} {\bibfnamefont
    {A.}~\bibnamefont {Ney}}, \bibinfo {author} {\bibfnamefont {H.}~\bibnamefont
    {Steiner}}, \bibinfo {author} {\bibfnamefont {V.~V.}\ \bibnamefont
    {Volobuev}}, \bibinfo {author} {\bibfnamefont {H.}~\bibnamefont {Groiss}},
    \bibinfo {author} {\bibfnamefont {M.}~\bibnamefont {Albu}}, \bibinfo {author}
    {\bibfnamefont {G.}~\bibnamefont {Kothleitner}}, \bibinfo {author}
    {\bibfnamefont {J.}~\bibnamefont {Michali{\v{c}}ka}}, \bibinfo {author}
    {\bibfnamefont {S.~A.}\ \bibnamefont {Khan}}, \bibinfo {author}
    {\bibfnamefont {J.}~\bibnamefont {Min{\'{a}}r}}, \bibinfo {author}
    {\bibfnamefont {H.}~\bibnamefont {Ebert}}, \bibinfo {author} {\bibfnamefont
    {G.}~\bibnamefont {Bauer}}, \bibinfo {author} {\bibfnamefont
    {F.}~\bibnamefont {Freyse}}, \bibinfo {author} {\bibfnamefont
    {A.}~\bibnamefont {Varykhalov}}, \bibinfo {author} {\bibfnamefont
    {O.}~\bibnamefont {Rader}}, \ and\ \bibinfo {author} {\bibfnamefont
    {G.}~\bibnamefont {Springholz}},\ }\href {\doibase 10.1038/s41586-019-1826-7}
    {\bibfield  {journal} {\bibinfo  {journal} {Nature}\ }\textbf {\bibinfo
    {volume} {576}},\ \bibinfo {pages} {423} (\bibinfo {year}
    {2019})}\BibitemShut {NoStop}%
  \bibitem [{\citenamefont {Otrokov}\ \emph {et~al.}(2019)\citenamefont
    {Otrokov}, \citenamefont {Klimovskikh}, \citenamefont {Bentmann},
    \citenamefont {Estyunin}, \citenamefont {Zeugner}, \citenamefont {Aliev},
    \citenamefont {Ga{\ss}}, \citenamefont {Wolter}, \citenamefont {Koroleva},
    \citenamefont {Shikin}, \citenamefont {Blanco-Rey}, \citenamefont {Hoffmann},
    \citenamefont {Rusinov}, \citenamefont {Vyazovskaya}, \citenamefont
    {Eremeev}, \citenamefont {Koroteev}, \citenamefont {Kuznetsov}, \citenamefont
    {Freyse}, \citenamefont {S{\'{a}}nchez-Barriga}, \citenamefont {Amiraslanov},
    \citenamefont {Babanly}, \citenamefont {Mamedov}, \citenamefont {Abdullayev},
    \citenamefont {Zverev}, \citenamefont {Alfonsov}, \citenamefont {Kataev},
    \citenamefont {Büchner}, \citenamefont {Schwier}, \citenamefont {Kumar},
    \citenamefont {Kimura}, \citenamefont {Petaccia}, \citenamefont {Santo},
    \citenamefont {Vidal}, \citenamefont {Schatz}, \citenamefont {Ki{\ss}ner},
    \citenamefont {Ünzelmann}, \citenamefont {Min}, \citenamefont {Moser},
    \citenamefont {Peixoto}, \citenamefont {Reinert}, \citenamefont {Ernst},
    \citenamefont {Echenique}, \citenamefont {Isaeva},\ and\ \citenamefont
    {Chulkov}}]{Otrokov_2019}%
    \BibitemOpen
    \bibfield  {author} {\bibinfo {author} {\bibfnamefont {M.~M.}\ \bibnamefont
    {Otrokov}}, \bibinfo {author} {\bibfnamefont {I.~I.}\ \bibnamefont
    {Klimovskikh}}, \bibinfo {author} {\bibfnamefont {H.}~\bibnamefont
    {Bentmann}}, \bibinfo {author} {\bibfnamefont {D.}~\bibnamefont {Estyunin}},
    \bibinfo {author} {\bibfnamefont {A.}~\bibnamefont {Zeugner}}, \bibinfo
    {author} {\bibfnamefont {Z.~S.}\ \bibnamefont {Aliev}}, \bibinfo {author}
    {\bibfnamefont {S.}~\bibnamefont {Ga{\ss}}}, \bibinfo {author} {\bibfnamefont
    {A.~U.~B.}\ \bibnamefont {Wolter}}, \bibinfo {author} {\bibfnamefont {A.~V.}\
    \bibnamefont {Koroleva}}, \bibinfo {author} {\bibfnamefont {A.~M.}\
    \bibnamefont {Shikin}}, \bibinfo {author} {\bibfnamefont {M.}~\bibnamefont
    {Blanco-Rey}}, \bibinfo {author} {\bibfnamefont {M.}~\bibnamefont
    {Hoffmann}}, \bibinfo {author} {\bibfnamefont {I.~P.}\ \bibnamefont
    {Rusinov}}, \bibinfo {author} {\bibfnamefont {A.~Y.}\ \bibnamefont
    {Vyazovskaya}}, \bibinfo {author} {\bibfnamefont {S.~V.}\ \bibnamefont
    {Eremeev}}, \bibinfo {author} {\bibfnamefont {Y.~M.}\ \bibnamefont
    {Koroteev}}, \bibinfo {author} {\bibfnamefont {V.~M.}\ \bibnamefont
    {Kuznetsov}}, \bibinfo {author} {\bibfnamefont {F.}~\bibnamefont {Freyse}},
    \bibinfo {author} {\bibfnamefont {J.}~\bibnamefont {S{\'{a}}nchez-Barriga}},
    \bibinfo {author} {\bibfnamefont {I.~R.}\ \bibnamefont {Amiraslanov}},
    \bibinfo {author} {\bibfnamefont {M.~B.}\ \bibnamefont {Babanly}}, \bibinfo
    {author} {\bibfnamefont {N.~T.}\ \bibnamefont {Mamedov}}, \bibinfo {author}
    {\bibfnamefont {N.~A.}\ \bibnamefont {Abdullayev}}, \bibinfo {author}
    {\bibfnamefont {V.~N.}\ \bibnamefont {Zverev}}, \bibinfo {author}
    {\bibfnamefont {A.}~\bibnamefont {Alfonsov}}, \bibinfo {author}
    {\bibfnamefont {V.}~\bibnamefont {Kataev}}, \bibinfo {author} {\bibfnamefont
    {B.}~\bibnamefont {Büchner}}, \bibinfo {author} {\bibfnamefont {E.~F.}\
    \bibnamefont {Schwier}}, \bibinfo {author} {\bibfnamefont {S.}~\bibnamefont
    {Kumar}}, \bibinfo {author} {\bibfnamefont {A.}~\bibnamefont {Kimura}},
    \bibinfo {author} {\bibfnamefont {L.}~\bibnamefont {Petaccia}}, \bibinfo
    {author} {\bibfnamefont {G.~D.}\ \bibnamefont {Santo}}, \bibinfo {author}
    {\bibfnamefont {R.~C.}\ \bibnamefont {Vidal}}, \bibinfo {author}
    {\bibfnamefont {S.}~\bibnamefont {Schatz}}, \bibinfo {author} {\bibfnamefont
    {K.}~\bibnamefont {Ki{\ss}ner}}, \bibinfo {author} {\bibfnamefont
    {M.}~\bibnamefont {Ünzelmann}}, \bibinfo {author} {\bibfnamefont {C.~H.}\
    \bibnamefont {Min}}, \bibinfo {author} {\bibfnamefont {S.}~\bibnamefont
    {Moser}}, \bibinfo {author} {\bibfnamefont {T.~R.~F.}\ \bibnamefont
    {Peixoto}}, \bibinfo {author} {\bibfnamefont {F.}~\bibnamefont {Reinert}},
    \bibinfo {author} {\bibfnamefont {A.}~\bibnamefont {Ernst}}, \bibinfo
    {author} {\bibfnamefont {P.~M.}\ \bibnamefont {Echenique}}, \bibinfo {author}
    {\bibfnamefont {A.}~\bibnamefont {Isaeva}}, \ and\ \bibinfo {author}
    {\bibfnamefont {E.~V.}\ \bibnamefont {Chulkov}},\ }\href {\doibase
    10.1038/s41586-019-1840-9} {\bibfield  {journal} {\bibinfo  {journal}
    {Nature}\ }\textbf {\bibinfo {volume} {576}},\ \bibinfo {pages} {416}
    (\bibinfo {year} {2019})}\BibitemShut {NoStop}%
  \bibitem [{\citenamefont {Zhang}\ \emph {et~al.}(2019)\citenamefont {Zhang},
    \citenamefont {Shi}, \citenamefont {Zhu}, \citenamefont {Xing}, \citenamefont
    {Zhang},\ and\ \citenamefont {Wang}}]{Zhang_2019}%
    \BibitemOpen
    \bibfield  {author} {\bibinfo {author} {\bibfnamefont {D.}~\bibnamefont
    {Zhang}}, \bibinfo {author} {\bibfnamefont {M.}~\bibnamefont {Shi}}, \bibinfo
    {author} {\bibfnamefont {T.}~\bibnamefont {Zhu}}, \bibinfo {author}
    {\bibfnamefont {D.}~\bibnamefont {Xing}}, \bibinfo {author} {\bibfnamefont
    {H.}~\bibnamefont {Zhang}}, \ and\ \bibinfo {author} {\bibfnamefont
    {J.}~\bibnamefont {Wang}},\ }\href {\doibase 10.1103/PhysRevLett.122.206401}
    {\bibfield  {journal} {\bibinfo  {journal} {Phys. Rev. Lett.}\ }\textbf
    {\bibinfo {volume} {122}},\ \bibinfo {pages} {206401} (\bibinfo {year}
    {2019})}\BibitemShut {NoStop}%
  \bibitem [{\citenamefont {Li}\ \emph {et~al.}(2019)\citenamefont {Li},
    \citenamefont {Li}, \citenamefont {Du}, \citenamefont {Wang}, \citenamefont
    {Gu}, \citenamefont {Zhang}, \citenamefont {He}, \citenamefont {Duan},\ and\
    \citenamefont {Xu}}]{Li_2019}%
    \BibitemOpen
    \bibfield  {author} {\bibinfo {author} {\bibfnamefont {J.}~\bibnamefont
    {Li}}, \bibinfo {author} {\bibfnamefont {Y.}~\bibnamefont {Li}}, \bibinfo
    {author} {\bibfnamefont {S.}~\bibnamefont {Du}}, \bibinfo {author}
    {\bibfnamefont {Z.}~\bibnamefont {Wang}}, \bibinfo {author} {\bibfnamefont
    {B.-L.}\ \bibnamefont {Gu}}, \bibinfo {author} {\bibfnamefont {S.-C.}\
    \bibnamefont {Zhang}}, \bibinfo {author} {\bibfnamefont {K.}~\bibnamefont
    {He}}, \bibinfo {author} {\bibfnamefont {W.}~\bibnamefont {Duan}}, \ and\
    \bibinfo {author} {\bibfnamefont {Y.}~\bibnamefont {Xu}},\ }\href {\doibase
    10.1126/sciadv.aaw5685} {\bibfield  {journal} {\bibinfo  {journal} {Science
    Advances}\ }\textbf {\bibinfo {volume} {5}} (\bibinfo {year} {2019}),\
    10.1126/sciadv.aaw5685}\BibitemShut {NoStop}%
  \bibitem [{\citenamefont {Zhao}\ \emph {et~al.}(2020)\citenamefont {Zhao},
    \citenamefont {Zhang}, \citenamefont {Mei}, \citenamefont {Zhou},
    \citenamefont {Yi}, \citenamefont {Zhang}, \citenamefont {Yu}, \citenamefont
    {Xiao}, \citenamefont {Wang}, \citenamefont {Samarth}, \citenamefont {Chan},
    \citenamefont {Liu},\ and\ \citenamefont {Chang}}]{Zhao_2020}%
    \BibitemOpen
    \bibfield  {author} {\bibinfo {author} {\bibfnamefont {Y.-F.}\ \bibnamefont
    {Zhao}}, \bibinfo {author} {\bibfnamefont {R.}~\bibnamefont {Zhang}},
    \bibinfo {author} {\bibfnamefont {R.}~\bibnamefont {Mei}}, \bibinfo {author}
    {\bibfnamefont {L.-J.}\ \bibnamefont {Zhou}}, \bibinfo {author}
    {\bibfnamefont {H.}~\bibnamefont {Yi}}, \bibinfo {author} {\bibfnamefont
    {Y.-Q.}\ \bibnamefont {Zhang}}, \bibinfo {author} {\bibfnamefont
    {J.}~\bibnamefont {Yu}}, \bibinfo {author} {\bibfnamefont {R.}~\bibnamefont
    {Xiao}}, \bibinfo {author} {\bibfnamefont {K.}~\bibnamefont {Wang}}, \bibinfo
    {author} {\bibfnamefont {N.}~\bibnamefont {Samarth}}, \bibinfo {author}
    {\bibfnamefont {M.~H.~W.}\ \bibnamefont {Chan}}, \bibinfo {author}
    {\bibfnamefont {C.-X.}\ \bibnamefont {Liu}}, \ and\ \bibinfo {author}
    {\bibfnamefont {C.-Z.}\ \bibnamefont {Chang}},\ }\href {\doibase
    10.1038/s41586-020-3020-3} {\bibfield  {journal} {\bibinfo  {journal}
    {Nature}\ }\textbf {\bibinfo {volume} {588}},\ \bibinfo {pages} {419}
    (\bibinfo {year} {2020})}\BibitemShut {NoStop}%
  \bibitem [{\citenamefont {Zhao}\ \emph {et~al.}(2021)\citenamefont {Zhao},
    \citenamefont {Zhang}, \citenamefont {Zhou}, \citenamefont {Mei},
    \citenamefont {Yan}, \citenamefont {Chan}, \citenamefont {Liu},\ and\
    \citenamefont {Chang}}]{Zhao_2021}%
    \BibitemOpen
    \bibfield  {author} {\bibinfo {author} {\bibfnamefont {Y.-F.}\ \bibnamefont
    {Zhao}}, \bibinfo {author} {\bibfnamefont {R.}~\bibnamefont {Zhang}},
    \bibinfo {author} {\bibfnamefont {L.-J.}\ \bibnamefont {Zhou}}, \bibinfo
    {author} {\bibfnamefont {R.}~\bibnamefont {Mei}}, \bibinfo {author}
    {\bibfnamefont {Z.-J.}\ \bibnamefont {Yan}}, \bibinfo {author} {\bibfnamefont
    {M.~H.~W.}\ \bibnamefont {Chan}}, \bibinfo {author} {\bibfnamefont {C.-X.}\
    \bibnamefont {Liu}}, \ and\ \bibinfo {author} {\bibfnamefont {C.-Z.}\
    \bibnamefont {Chang}},\ }\href {\doibase 10.48550/ARXIV.2109.11382} {\enquote
    {\bibinfo {title} {Zero magnetic field plateau phase transition in higher
    chern number quantum anomalous hall insulators},}\ } (\bibinfo {year}
    {2021})\BibitemShut {NoStop}%
  \bibitem [{\citenamefont {Deng}\ \emph {et~al.}(2020)\citenamefont {Deng},
    \citenamefont {Yu}, \citenamefont {Shi}, \citenamefont {Guo}, \citenamefont
    {Xu}, \citenamefont {Wang}, \citenamefont {Chen},\ and\ \citenamefont
    {Zhang}}]{Deng_2020}%
    \BibitemOpen
    \bibfield  {author} {\bibinfo {author} {\bibfnamefont {Y.}~\bibnamefont
    {Deng}}, \bibinfo {author} {\bibfnamefont {Y.}~\bibnamefont {Yu}}, \bibinfo
    {author} {\bibfnamefont {M.~Z.}\ \bibnamefont {Shi}}, \bibinfo {author}
    {\bibfnamefont {Z.}~\bibnamefont {Guo}}, \bibinfo {author} {\bibfnamefont
    {Z.}~\bibnamefont {Xu}}, \bibinfo {author} {\bibfnamefont {J.}~\bibnamefont
    {Wang}}, \bibinfo {author} {\bibfnamefont {X.~H.}\ \bibnamefont {Chen}}, \
    and\ \bibinfo {author} {\bibfnamefont {Y.}~\bibnamefont {Zhang}},\ }\href
    {\doibase 10.1126/science.aax8156} {\bibfield  {journal} {\bibinfo  {journal}
    {Science}\ }\textbf {\bibinfo {volume} {367}},\ \bibinfo {pages} {895}
    (\bibinfo {year} {2020})}\BibitemShut {NoStop}%
  \bibitem [{\citenamefont {Ge}\ \emph {et~al.}(2020{\natexlab{a}})\citenamefont
    {Ge}, \citenamefont {Liu}, \citenamefont {Li}, \citenamefont {Li},
    \citenamefont {Luo}, \citenamefont {Wu}, \citenamefont {Xu},\ and\
    \citenamefont {Wang}}]{GeWangJ}%
    \BibitemOpen
    \bibfield  {author} {\bibinfo {author} {\bibfnamefont {J.}~\bibnamefont
    {Ge}}, \bibinfo {author} {\bibfnamefont {Y.}~\bibnamefont {Liu}}, \bibinfo
    {author} {\bibfnamefont {J.}~\bibnamefont {Li}}, \bibinfo {author}
    {\bibfnamefont {H.}~\bibnamefont {Li}}, \bibinfo {author} {\bibfnamefont
    {T.}~\bibnamefont {Luo}}, \bibinfo {author} {\bibfnamefont {Y.}~\bibnamefont
    {Wu}}, \bibinfo {author} {\bibfnamefont {Y.}~\bibnamefont {Xu}}, \ and\
    \bibinfo {author} {\bibfnamefont {J.}~\bibnamefont {Wang}},\ }\href {\doibase
    10.1093/nsr/nwaa089} {\bibfield  {journal} {\bibinfo  {journal} {National
    Science Review}\ }\textbf {\bibinfo {volume} {7}},\ \bibinfo {pages} {1280}
    (\bibinfo {year} {2020}{\natexlab{a}})}\BibitemShut {NoStop}%
  \bibitem [{\citenamefont {Liu}\ \emph {et~al.}(2020)\citenamefont {Liu},
    \citenamefont {Wang}, \citenamefont {Li}, \citenamefont {Wu}, \citenamefont
    {Li}, \citenamefont {Li}, \citenamefont {He}, \citenamefont {Xu},
    \citenamefont {Zhang},\ and\ \citenamefont {Wang}}]{Wang.Y}%
    \BibitemOpen
    \bibfield  {author} {\bibinfo {author} {\bibfnamefont {C.}~\bibnamefont
    {Liu}}, \bibinfo {author} {\bibfnamefont {Y.}~\bibnamefont {Wang}}, \bibinfo
    {author} {\bibfnamefont {H.}~\bibnamefont {Li}}, \bibinfo {author}
    {\bibfnamefont {Y.}~\bibnamefont {Wu}}, \bibinfo {author} {\bibfnamefont
    {Y.}~\bibnamefont {Li}}, \bibinfo {author} {\bibfnamefont {J.}~\bibnamefont
    {Li}}, \bibinfo {author} {\bibfnamefont {K.}~\bibnamefont {He}}, \bibinfo
    {author} {\bibfnamefont {Y.}~\bibnamefont {Xu}}, \bibinfo {author}
    {\bibfnamefont {J.}~\bibnamefont {Zhang}}, \ and\ \bibinfo {author}
    {\bibfnamefont {Y.}~\bibnamefont {Wang}},\ }\href {\doibase
    10.1038/s41563-019-0573-3} {\bibfield  {journal} {\bibinfo  {journal} {Nature
    Materials}\ }\textbf {\bibinfo {volume} {19}},\ \bibinfo {pages} {522}
    (\bibinfo {year} {2020})}\BibitemShut {NoStop}%
  \bibitem [{\citenamefont {Zhang}\ \emph {et~al.}(2020)\citenamefont {Zhang},
    \citenamefont {Wu},\ and\ \citenamefont {Das~Sarma}}]{Das}%
    \BibitemOpen
    \bibfield  {author} {\bibinfo {author} {\bibfnamefont {R.-X.}\ \bibnamefont
    {Zhang}}, \bibinfo {author} {\bibfnamefont {F.}~\bibnamefont {Wu}}, \ and\
    \bibinfo {author} {\bibfnamefont {S.}~\bibnamefont {Das~Sarma}},\ }\href
    {\doibase 10.1103/PhysRevLett.124.136407} {\bibfield  {journal} {\bibinfo
    {journal} {Phys. Rev. Lett.}\ }\textbf {\bibinfo {volume} {124}},\ \bibinfo
    {pages} {136407} (\bibinfo {year} {2020})}\BibitemShut {NoStop}%
  \bibitem [{\citenamefont {Lian}\ \emph {et~al.}(2020)\citenamefont {Lian},
    \citenamefont {Liu}, \citenamefont {Zhang},\ and\ \citenamefont
    {Wang}}]{Liantwist}%
    \BibitemOpen
    \bibfield  {author} {\bibinfo {author} {\bibfnamefont {B.}~\bibnamefont
    {Lian}}, \bibinfo {author} {\bibfnamefont {Z.}~\bibnamefont {Liu}}, \bibinfo
    {author} {\bibfnamefont {Y.}~\bibnamefont {Zhang}}, \ and\ \bibinfo {author}
    {\bibfnamefont {J.}~\bibnamefont {Wang}},\ }\href {\doibase
    10.1103/PhysRevLett.124.126402} {\bibfield  {journal} {\bibinfo  {journal}
    {Phys Rev Lett}\ }\textbf {\bibinfo {volume} {124}},\ \bibinfo {pages}
    {126402} (\bibinfo {year} {2020})}\BibitemShut {NoStop}%
  \bibitem [{\citenamefont {Sun}\ \emph {et~al.}(2019)\citenamefont {Sun},
    \citenamefont {Xia}, \citenamefont {Chen}, \citenamefont {Zhang},
    \citenamefont {Liu}, \citenamefont {Yao}, \citenamefont {Tang}, \citenamefont
    {Zhao}, \citenamefont {Xu},\ and\ \citenamefont {Liu}}]{LiuQH}%
    \BibitemOpen
    \bibfield  {author} {\bibinfo {author} {\bibfnamefont {H.}~\bibnamefont
    {Sun}}, \bibinfo {author} {\bibfnamefont {B.}~\bibnamefont {Xia}}, \bibinfo
    {author} {\bibfnamefont {Z.}~\bibnamefont {Chen}}, \bibinfo {author}
    {\bibfnamefont {Y.}~\bibnamefont {Zhang}}, \bibinfo {author} {\bibfnamefont
    {P.}~\bibnamefont {Liu}}, \bibinfo {author} {\bibfnamefont {Q.}~\bibnamefont
    {Yao}}, \bibinfo {author} {\bibfnamefont {H.}~\bibnamefont {Tang}}, \bibinfo
    {author} {\bibfnamefont {Y.}~\bibnamefont {Zhao}}, \bibinfo {author}
    {\bibfnamefont {H.}~\bibnamefont {Xu}}, \ and\ \bibinfo {author}
    {\bibfnamefont {Q.}~\bibnamefont {Liu}},\ }\href {\doibase
    10.1103/PhysRevLett.123.096401} {\bibfield  {journal} {\bibinfo  {journal}
    {Phys Rev Lett}\ }\textbf {\bibinfo {volume} {123}},\ \bibinfo {pages}
    {096401} (\bibinfo {year} {2019})}\BibitemShut {NoStop}%
  \bibitem [{\citenamefont {Li}\ \emph {et~al.}(2021)\citenamefont {Li},
    \citenamefont {Jiang}, \citenamefont {Chen},\ and\ \citenamefont
    {Xie}}]{LHLlocallaxion}%
    \BibitemOpen
    \bibfield  {author} {\bibinfo {author} {\bibfnamefont {H.}~\bibnamefont
    {Li}}, \bibinfo {author} {\bibfnamefont {H.}~\bibnamefont {Jiang}}, \bibinfo
    {author} {\bibfnamefont {C.~Z.}\ \bibnamefont {Chen}}, \ and\ \bibinfo
    {author} {\bibfnamefont {X.~C.}\ \bibnamefont {Xie}},\ }\href {\doibase
    10.1103/PhysRevLett.126.156601} {\bibfield  {journal} {\bibinfo  {journal}
    {Phys Rev Lett}\ }\textbf {\bibinfo {volume} {126}},\ \bibinfo {pages}
    {156601} (\bibinfo {year} {2021})}\BibitemShut {NoStop}%
  \bibitem [{\citenamefont {Gao}\ \emph {et~al.}(2021)\citenamefont {Gao},
    \citenamefont {Liu}, \citenamefont {Hu}, \citenamefont {Qiu}, \citenamefont
    {Tzschaschel}, \citenamefont {Ghosh}, \citenamefont {Ho}, \citenamefont
    {Berube}, \citenamefont {Chen}, \citenamefont {Sun}, \citenamefont {Zhang},
    \citenamefont {Zhang}, \citenamefont {Wang}, \citenamefont {Wang},
    \citenamefont {Huang}, \citenamefont {Felser}, \citenamefont {Agarwal},
    \citenamefont {Ding}, \citenamefont {Tien}, \citenamefont {Akey},
    \citenamefont {Gardener}, \citenamefont {Singh}, \citenamefont {Watanabe},
    \citenamefont {Taniguchi}, \citenamefont {Burch}, \citenamefont {Bell},
    \citenamefont {Zhou}, \citenamefont {Gao}, \citenamefont {Lu}, \citenamefont
    {Bansil}, \citenamefont {Lin}, \citenamefont {Chang}, \citenamefont {Fu},
    \citenamefont {Ma}, \citenamefont {Ni},\ and\ \citenamefont {Xu}}]{LHE}%
    \BibitemOpen
    \bibfield  {author} {\bibinfo {author} {\bibfnamefont {A.}~\bibnamefont
    {Gao}}, \bibinfo {author} {\bibfnamefont {Y.~F.}\ \bibnamefont {Liu}},
    \bibinfo {author} {\bibfnamefont {C.}~\bibnamefont {Hu}}, \bibinfo {author}
    {\bibfnamefont {J.~X.}\ \bibnamefont {Qiu}}, \bibinfo {author} {\bibfnamefont
    {C.}~\bibnamefont {Tzschaschel}}, \bibinfo {author} {\bibfnamefont
    {B.}~\bibnamefont {Ghosh}}, \bibinfo {author} {\bibfnamefont {S.~C.}\
    \bibnamefont {Ho}}, \bibinfo {author} {\bibfnamefont {D.}~\bibnamefont
    {Berube}}, \bibinfo {author} {\bibfnamefont {R.}~\bibnamefont {Chen}},
    \bibinfo {author} {\bibfnamefont {H.}~\bibnamefont {Sun}}, \bibinfo {author}
    {\bibfnamefont {Z.}~\bibnamefont {Zhang}}, \bibinfo {author} {\bibfnamefont
    {X.~Y.}\ \bibnamefont {Zhang}}, \bibinfo {author} {\bibfnamefont {Y.~X.}\
    \bibnamefont {Wang}}, \bibinfo {author} {\bibfnamefont {N.}~\bibnamefont
    {Wang}}, \bibinfo {author} {\bibfnamefont {Z.}~\bibnamefont {Huang}},
    \bibinfo {author} {\bibfnamefont {C.}~\bibnamefont {Felser}}, \bibinfo
    {author} {\bibfnamefont {A.}~\bibnamefont {Agarwal}}, \bibinfo {author}
    {\bibfnamefont {T.}~\bibnamefont {Ding}}, \bibinfo {author} {\bibfnamefont
    {H.~J.}\ \bibnamefont {Tien}}, \bibinfo {author} {\bibfnamefont
    {A.}~\bibnamefont {Akey}}, \bibinfo {author} {\bibfnamefont {J.}~\bibnamefont
    {Gardener}}, \bibinfo {author} {\bibfnamefont {B.}~\bibnamefont {Singh}},
    \bibinfo {author} {\bibfnamefont {K.}~\bibnamefont {Watanabe}}, \bibinfo
    {author} {\bibfnamefont {T.}~\bibnamefont {Taniguchi}}, \bibinfo {author}
    {\bibfnamefont {K.~S.}\ \bibnamefont {Burch}}, \bibinfo {author}
    {\bibfnamefont {D.~C.}\ \bibnamefont {Bell}}, \bibinfo {author}
    {\bibfnamefont {B.~B.}\ \bibnamefont {Zhou}}, \bibinfo {author}
    {\bibfnamefont {W.}~\bibnamefont {Gao}}, \bibinfo {author} {\bibfnamefont
    {H.~Z.}\ \bibnamefont {Lu}}, \bibinfo {author} {\bibfnamefont
    {A.}~\bibnamefont {Bansil}}, \bibinfo {author} {\bibfnamefont
    {H.}~\bibnamefont {Lin}}, \bibinfo {author} {\bibfnamefont {T.~R.}\
    \bibnamefont {Chang}}, \bibinfo {author} {\bibfnamefont {L.}~\bibnamefont
    {Fu}}, \bibinfo {author} {\bibfnamefont {Q.}~\bibnamefont {Ma}}, \bibinfo
    {author} {\bibfnamefont {N.}~\bibnamefont {Ni}}, \ and\ \bibinfo {author}
    {\bibfnamefont {S.~Y.}\ \bibnamefont {Xu}},\ }\href {\doibase
    10.1038/s41586-021-03679-w} {\bibfield  {journal} {\bibinfo  {journal}
    {Nature}\ }\textbf {\bibinfo {volume} {595}},\ \bibinfo {pages} {521}
    (\bibinfo {year} {2021})}\BibitemShut {NoStop}%
  \bibitem [{sup()}]{supp}%
    \BibitemOpen
    \href@noop {} {}\bibinfo {note} {See the supplementary materials for the
    layer Hall effect in the weak disorder strength, more discussions on the
    Berry curvature of the topological magnet $\mathrm{MnBi_2Te_4}$ and the
    numerical methods. Additionally, the experimental characterization of the
    layer-locked QAHE edge mode.}\BibitemShut {Stop}%
  \bibitem [{\citenamefont {Prodan}(2009)}]{Prodan_2009}%
    \BibitemOpen
    \bibfield  {author} {\bibinfo {author} {\bibfnamefont {E.}~\bibnamefont
    {Prodan}},\ }\href {\doibase 10.1103/PhysRevB.80.125327} {\bibfield
    {journal} {\bibinfo  {journal} {Phys. Rev. B}\ }\textbf {\bibinfo {volume}
    {80}},\ \bibinfo {pages} {125327} (\bibinfo {year} {2009})}\BibitemShut
    {NoStop}%
  \bibitem [{\citenamefont {Prodan}(2011)}]{Prodan_2011}%
    \BibitemOpen
    \bibfield  {author} {\bibinfo {author} {\bibfnamefont {E.}~\bibnamefont
    {Prodan}},\ }\href {\doibase 10.1088/1751-8113/44/11/113001} {\bibfield
    {journal} {\bibinfo  {journal} {Journal of Physics A: Mathematical and
    Theoretical}\ }\textbf {\bibinfo {volume} {44}},\ \bibinfo {pages} {113001}
    (\bibinfo {year} {2011})}\BibitemShut {NoStop}%
  \bibitem [{\citenamefont {Chen}\ \emph {et~al.}(2021)\citenamefont {Chen},
    \citenamefont {Qi}, \citenamefont {Xu},\ and\ \citenamefont
    {Xie}}]{CCZ2021evolution}%
    \BibitemOpen
    \bibfield  {author} {\bibinfo {author} {\bibfnamefont {C.-Z.}\ \bibnamefont
    {Chen}}, \bibinfo {author} {\bibfnamefont {J.}~\bibnamefont {Qi}}, \bibinfo
    {author} {\bibfnamefont {D.-H.}\ \bibnamefont {Xu}}, \ and\ \bibinfo {author}
    {\bibfnamefont {X.}~\bibnamefont {Xie}},\ }\href {\doibase
    10.1007/s11433-021-1774-1} {\bibfield  {journal} {\bibinfo  {journal}
    {Science China Physics, Mechanics \& Astronomy}\ }\textbf {\bibinfo {volume}
    {64}},\ \bibinfo {pages} {1} (\bibinfo {year} {2021})}\BibitemShut {NoStop}%
  \bibitem [{\citenamefont {Jiang}\ \emph {et~al.}(2009)\citenamefont {Jiang},
    \citenamefont {Wang}, \citenamefont {Sun},\ and\ \citenamefont
    {Xie}}]{Jiang2009}%
    \BibitemOpen
    \bibfield  {author} {\bibinfo {author} {\bibfnamefont {H.}~\bibnamefont
    {Jiang}}, \bibinfo {author} {\bibfnamefont {L.}~\bibnamefont {Wang}},
    \bibinfo {author} {\bibfnamefont {Q.-f.}\ \bibnamefont {Sun}}, \ and\
    \bibinfo {author} {\bibfnamefont {X.~C.}\ \bibnamefont {Xie}},\ }\href
    {\doibase 10.1103/PhysRevB.80.165316} {\bibfield  {journal} {\bibinfo
    {journal} {Phys. Rev. B}\ }\textbf {\bibinfo {volume} {80}},\ \bibinfo
    {pages} {165316} (\bibinfo {year} {2009})}\BibitemShut {NoStop}%
  \bibitem [{\citenamefont {Zhang}\ \emph {et~al.}(2012)\citenamefont {Zhang},
    \citenamefont {Chu}, \citenamefont {Zhang},\ and\ \citenamefont
    {Shen}}]{ZYYgap}%
    \BibitemOpen
    \bibfield  {author} {\bibinfo {author} {\bibfnamefont {Y.-Y.}\ \bibnamefont
    {Zhang}}, \bibinfo {author} {\bibfnamefont {R.-L.}\ \bibnamefont {Chu}},
    \bibinfo {author} {\bibfnamefont {F.-C.}\ \bibnamefont {Zhang}}, \ and\
    \bibinfo {author} {\bibfnamefont {S.-Q.}\ \bibnamefont {Shen}},\ }\href
    {\doibase 10.1103/PhysRevB.85.035107} {\bibfield  {journal} {\bibinfo
    {journal} {Phys. Rev. B}\ }\textbf {\bibinfo {volume} {85}},\ \bibinfo
    {pages} {035107} (\bibinfo {year} {2012})}\BibitemShut {NoStop}%
  \bibitem [{\citenamefont {Zhang}\ and\ \citenamefont {Shen}(2013)}]{ZYYDOSTAI}%
    \BibitemOpen
    \bibfield  {author} {\bibinfo {author} {\bibfnamefont {Y.-Y.}\ \bibnamefont
    {Zhang}}\ and\ \bibinfo {author} {\bibfnamefont {S.-Q.}\ \bibnamefont
    {Shen}},\ }\href {\doibase 10.1103/PhysRevB.88.195145} {\bibfield  {journal}
    {\bibinfo  {journal} {Phys. Rev. B}\ }\textbf {\bibinfo {volume} {88}},\
    \bibinfo {pages} {195145} (\bibinfo {year} {2013})}\BibitemShut {NoStop}%
  \bibitem [{\citenamefont {Dobrosavljevi{\'{c}}}\ \emph
    {et~al.}(2003)\citenamefont {Dobrosavljevi{\'{c}}}, \citenamefont {Pastor},\
    and\ \citenamefont {Nikoli{\'{c}}}}]{Dobrosavljevi__2003}%
    \BibitemOpen
    \bibfield  {author} {\bibinfo {author} {\bibfnamefont {V.}~\bibnamefont
    {Dobrosavljevi{\'{c}}}}, \bibinfo {author} {\bibfnamefont {A.~A.}\
    \bibnamefont {Pastor}}, \ and\ \bibinfo {author} {\bibfnamefont {B.~K.}\
    \bibnamefont {Nikoli{\'{c}}}},\ }\href {\doibase 10.1209/epl/i2003-00364-5}
    {\bibfield  {journal} {\bibinfo  {journal} {Europhysics Letters ({EPL})}\
    }\textbf {\bibinfo {volume} {62}},\ \bibinfo {pages} {76} (\bibinfo {year}
    {2003})}\BibitemShut {NoStop}%
  \bibitem [{\citenamefont {Schubert}\ \emph {et~al.}(2010)\citenamefont
    {Schubert}, \citenamefont {Schleede}, \citenamefont {Byczuk}, \citenamefont
    {Fehske},\ and\ \citenamefont {Vollhardt}}]{Schubert}%
    \BibitemOpen
    \bibfield  {author} {\bibinfo {author} {\bibfnamefont {G.}~\bibnamefont
    {Schubert}}, \bibinfo {author} {\bibfnamefont {J.}~\bibnamefont {Schleede}},
    \bibinfo {author} {\bibfnamefont {K.}~\bibnamefont {Byczuk}}, \bibinfo
    {author} {\bibfnamefont {H.}~\bibnamefont {Fehske}}, \ and\ \bibinfo {author}
    {\bibfnamefont {D.}~\bibnamefont {Vollhardt}},\ }\href {\doibase
    10.1103/PhysRevB.81.155106} {\bibfield  {journal} {\bibinfo  {journal} {Phys.
    Rev. B}\ }\textbf {\bibinfo {volume} {81}},\ \bibinfo {pages} {155106}
    (\bibinfo {year} {2010})}\BibitemShut {NoStop}%
  \bibitem [{\citenamefont {Janssen}(1998)}]{JANSSEN19981}%
    \BibitemOpen
    \bibfield  {author} {\bibinfo {author} {\bibfnamefont {M.}~\bibnamefont
    {Janssen}},\ }\href {\doibase https://doi.org/10.1016/S0370-1573(97)00050-1}
    {\bibfield  {journal} {\bibinfo  {journal} {Physics Reports}\ }\textbf
    {\bibinfo {volume} {295}},\ \bibinfo {pages} {1} (\bibinfo {year}
    {1998})}\BibitemShut {NoStop}%
  \bibitem [{\citenamefont {Ge}\ \emph {et~al.}(2020{\natexlab{b}})\citenamefont
    {Ge}, \citenamefont {Liu}, \citenamefont {Li}, \citenamefont {Li},
    \citenamefont {Luo}, \citenamefont {Wu}, \citenamefont {Xu},\ and\
    \citenamefont {Wang}}]{Ge_2020}%
    \BibitemOpen
    \bibfield  {author} {\bibinfo {author} {\bibfnamefont {J.}~\bibnamefont
    {Ge}}, \bibinfo {author} {\bibfnamefont {Y.}~\bibnamefont {Liu}}, \bibinfo
    {author} {\bibfnamefont {J.}~\bibnamefont {Li}}, \bibinfo {author}
    {\bibfnamefont {H.}~\bibnamefont {Li}}, \bibinfo {author} {\bibfnamefont
    {T.}~\bibnamefont {Luo}}, \bibinfo {author} {\bibfnamefont {Y.}~\bibnamefont
    {Wu}}, \bibinfo {author} {\bibfnamefont {Y.}~\bibnamefont {Xu}}, \ and\
    \bibinfo {author} {\bibfnamefont {J.}~\bibnamefont {Wang}},\ }\href {\doibase
    10.1093/nsr/nwaa089} {\bibfield  {journal} {\bibinfo  {journal} {National
    Science Review}\ }\textbf {\bibinfo {volume} {7}},\ \bibinfo {pages} {1280}
    (\bibinfo {year} {2020}{\natexlab{b}})}\BibitemShut {NoStop}%
  \end{thebibliography}
%

\end{document}


\title{Supplementary Materials for ``Quantum Anomalous Layer Hall Effect in the Topological Magnet MnBi$_2$Te$_4$''}
\author{Wen-Bo Dai}
\affiliation{International Center for Quantum Materials, School of Physics, Peking University, Beijing 100871, China}
\affiliation{Beijing Academy of Quantum Information Sciences, Beijing 100193, China}
\author{Hailong Li}
\affiliation{International Center for Quantum Materials, School of Physics, Peking University, Beijing 100871, China}
\author{Dong-Hui Xu}
\affiliation{Department of Physics, Chongqing University, Chongqing 400044, China}
\email{czchen@suda.edu.cn}
\author{Chui-Zhen Chen}
\email{czchen@suda.edu.cn}
\affiliation{School of Physical Science and Technology, Soochow University, Suzhou 215006, China}
\affiliation{Institute for Advanced Study, Soochow University, Suzhou 215006, China}
\author{X. C. Xie}
\email{xcxie@pku.edu.cn}
\affiliation{International Center for Quantum Materials, School of Physics, Peking University, Beijing 100871, China}
\affiliation{Collaborative Innovation Center of Quantum Matter, Beijing 100871, China}
\affiliation{CAS Center for Excellence in Topological Quantum Computation, University of Chinese Academy of Sciences, Beijing 100190, China}
\affiliation{Beijing Academy of Quantum Information Sciences, Beijing 100193, China}
\date{\today }

\maketitle
\section{The Layer Hall effect in the weak disordered $\mathrm{MnBi_{2}Te_{4}}$}
\begin{figure}[htbp]
	\includegraphics[width=6.5in]{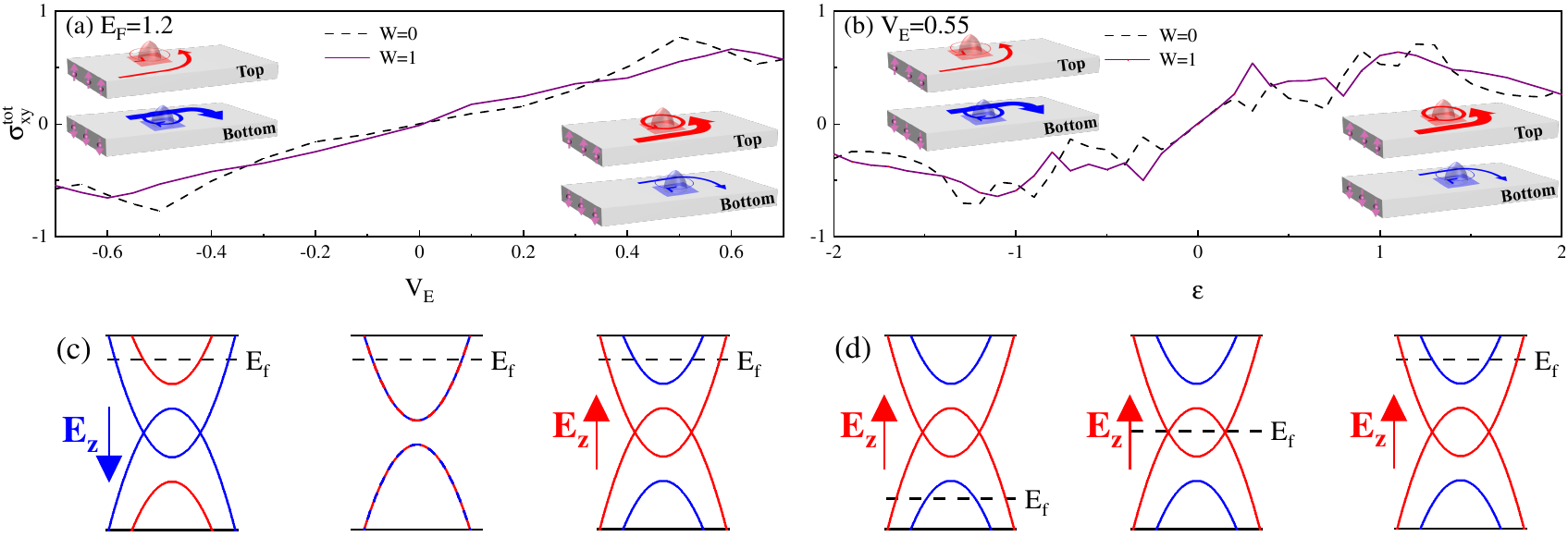}
	\caption{(a) The total Chern number $C_{tot}$ as a function of electric field $V_E$ with Fermi energy $E_f=1.2$.
	(b) The total Chern number $C_{tot}$ as a function of electric field $E_f$ with Fermi energy $V_E=0.55$.
	(c)-(d) The schematic of the evolution of the energy band with varying $V_E$ or $E_f$ corresponding to (a)-(b)
	}
	\label{figS1}
\end{figure}

In this section, we show that the layer Hall effect (LHE) exists in the even-layered antiferromagnetic (AFM) $\mathrm{MnBi_{2}Te_{4}}$ in the clean limit or with the weak disorder strength.
In Fig.~\ref{figS1}(a), we plot the Hall conductance as a function of the electric-field strength $V_E$ with $E_f=1.2$.  One can see that the anomalous Hall effect (AHE) with the positive or negative Hall conductance is induced by the top or bottom layer-polarized net Berry curvature under an upward or downward vertical electric field. 
Alternatively, in Fig.~\ref{figS1}(b), we find that tuning the Fermi energy  $E_f$ can induce a transition from the top layer-polarized AHE with the positive net  Berry curvature
to the bottom layer-polarized AHE  with negative net Berry curvature.
Note that the LHE is sustainable to weak disorder at  $W=1$([see the solid line in Fig.\ref{figS1}(a-b)]).
 These results are qualitatively consistent with those discovered in the recent experiment.

The LHE can be understood by the band theory in the clean limit as follows.
Generally, as shown in Fig.~\ref{figS1}(c-d), there are one pair of conduction band with positive Berry curvature (red line) and valence band with negative Berry curvature (blue line) for the top layer,
while the other pair of degenerate parity-time-reversal bands for the bottom layer are of opposite Berry curvature. To be specific, in Fig.~\ref{figS1}(c), the bands of the top and the bottom layer are degenerate
with the opposite Berry curvature in the absence of the electric field. Then, by applying an upward (downward) electric field,  the band degeneracy is lifted, giving rise to a 
 net positive (negative) Berry curvature or Hall conductance in the whole system  [see Fig.~\ref{figS1}(c)].
 Similarly,  in Fig.~\ref{figS1}(d),  when the system starts from a fixed upward electric field ($V_E=1.2$),
tuning the Fermi energy  $E_f$ across the neutral point ($E_f=0$) can create a crossover from net negative bottom-layer-locked Berry curvature to  net negative top-layer-locked Berry curvature.

\section{$\mathcal{P} \mathcal{T} $ Symmetry and the Layer-Dependent Berry curvature}
\begin{figure}[htbp]
    \includegraphics[width=6.5in]{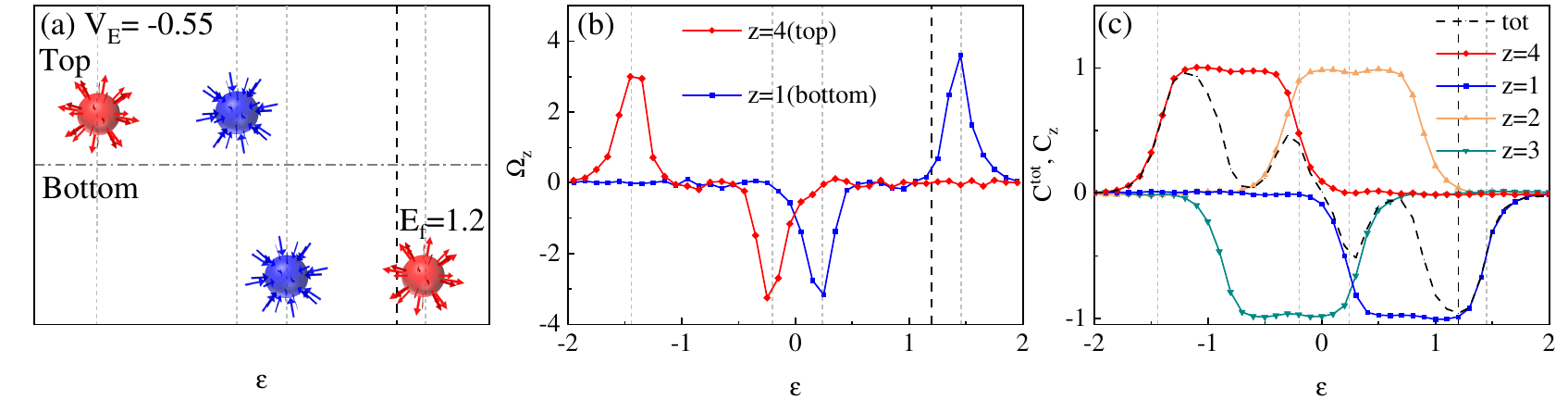}
    \caption{(color online) (a)Schematic plots of the the Berry-curvature monopoles in AFM MnBi$_2$Te$_4$ when the $E$ field strength $V_E=-0.55$
	Corresponding, (b) shows layer-dependent Berry curvature versus the energy $\epsilon$,
	and (c) shows the layer-dependent and the total Chern numbers. The Fermi energy $E_f$ = 1.2 is fixed with the system size $48 \times  48 \times 4$.}
	\label{figS2}
  \end{figure}

 In Fig.~2 of the main text, we have quantitatively elucidated how an upward electric field
creates a Berry-curvature monopole on the bottom layer. Due to the parity-time ( $\mathcal{P} \mathcal{T} $) symmetry,  the upward electric field can drive a positive monopole in the top layer to above the Fermi energy.
Here, we consider the situation with a downward electric field. In Fig.~\ref{figS2},  we evaluate the Chern number and the Berry curvature as a function energy $\varepsilon$, and then schematically plot the Berry curvature monopoles correspondingly in the energy space. Comparing to the Fig.~2 of the main text, one can find that the downward electric field can trigger a positive monopole on the bottom layer to above the Fermi energy.

Next, we will prove that, due to the $\mathcal{P} \mathcal{T} $ symmetry of the even-layered AFM $\mathrm{MnBi_{2}Te_{4}}$ without external fields (see $H$ in the main text), the reverse vertical electric field can polarize the Berry curvature of the opposite value on the opposite ($+z$ and $-z$) layers, i.e.
\begin{eqnarray}\label{eqS1}
	\begin{array}{c}
		C_z(E_f,-V_E)=-C_{\mathcal{P}z}(E_f,+V_E) \\
		\\
	  \Omega_z(\varepsilon,-V_E )=-\Omega_{\mathcal{P}z}(\varepsilon,+V_E )
	\end{array}
  \end{eqnarray}
For simplicity, we denote $\mathcal{P} \mathcal{T}=\Theta $ and it is straight forward to verify that 
\begin{eqnarray}\label{eqS2}
	\begin{array}{c}
		 H (-V_E)\cdot\Theta  =\Theta \cdot H (V_E)\\
		\\
	\end{array}
  \end{eqnarray}
  where $V_E$ is the $E$-field strength.
If $|\epsilon(V_E)\rangle  $ is an eigenstate  of $ H^- (V_E)$  with energy $\epsilon$, its space-time $\mathcal{P} \mathcal{T} $ partner $\Theta|\epsilon(V_E)\rangle  $
is an eigenstate of $ H^- (-V_E,)$ with energy $\epsilon$. Thus, we can conclude that
\begin{eqnarray}\label{eqS3}
	\begin{array}{c}
		P_{Ef}(-V_E) = \Theta P_{Ef}(V_E)\Theta^\dagger\\
	\end{array}
  \end{eqnarray}
If we denote $\left\lfloor \hat{x},P(V_E) \right\rfloor=K^{(x)}(V_E)$, the real-space noncommutative Kubo formula can be expressed as
\cite{Prodan_2009,Prodan_2011}
  \begin{eqnarray}\label{eqS4}
	C_z (Ef,V_E) &=& \sigma_{xy}(z,E_f,V_E)/(e^2/h)\nonumber\\
	&=&2\pi i \cdot \mathrm{Tr}\{ P_{E_f}(V_E ) [-i\lfloor \hat{x},P_{E_f}(V_E)\rfloor ,-i\lfloor \hat{y},P_{E_f}(V_E)\rfloor ]\}_z\nonumber\\
	&=&-2\pi i \cdot \mathrm{Tr}\{ P_{E_f}(V_E ) [K^{(x)}_{E_f}(V_E) ,K^{(y)}_{E_f}(V_E) ]\}_z
	\end{eqnarray}
We can express $K^{(x)}(-V_E)$ by using $P(-V_E)$ in Eq.~\ref{eqS3} as
\begin{eqnarray}\label{eqS5}
	K^{(x)}(-V_E) &=& \left\lfloor \hat{x},P(-V_E) \right\rfloor\nonumber\\
	&=&i\sum_{m=1}^Q\{C_m(e^{-im\hat{x}\Delta_i}P(-V_E)e^{im\hat{x}\Delta_i}-e^{im\hat{x}\Delta_i}P(-V_E)e^{-im\hat{x}\Delta_i})\}\nonumber\\
	&=&i\sum_{m=1}^Q\{C_m(e^{-im\hat{x}\Delta_i}\Theta P(V_E)\Theta^\dagger e^{im\hat{x}\Delta_i}-e^{im\hat{x}\Delta_i}\Theta P(V_E)\Theta^\dagger e^{-im\hat{x}\Delta_i})\}\nonumber\\
	&=&\Theta\{-i\sum_{m=1}^Q\{C_m(e^{im\hat{x}\Delta_i}P(V_E)e^{-im\hat{x}\Delta_i}-e^{-im\hat{x}\Delta_i} P(V_E)e^{im\hat{x}\Delta_i})\}\}\Theta^\dagger\nonumber\\
	&=& -\Theta K^{(-x)} _{V_E} \Theta^\dagger
	\end{eqnarray}
Replacing $V_E$ with $-V_E$ and substituting Eq.~\ref{eqS5} and Eq.~\ref{eqS3}  into Eq.~\ref{eqS4} we have
\begin{eqnarray}\label{eqS6}
	C_z (E_f,-V_E)&=&-2\pi i \cdot \mathrm{Tr}\{\Theta P_{E_f}(V_E ) \Theta^\dagger[-\Theta K^{(-x)}_{E_f}(V_E) \Theta^\dagger ,-\Theta K^{(-y)}_{E_f}(V_E) \Theta^\dagger ]\}_z\nonumber\\
	&=& -2\pi i \sum_{n \in z\& l, k\in all }\{P_{nl}^{\Theta}(V_E ) (K_{l,k}^{(-x)\Theta}(V_E) K_{k,n}^{(-y)\Theta}(V_E)-K_{l,k}^{(-y)\Theta}(V_E) K_{k,n}^{(-x)\Theta}(V_E))\}
	\end{eqnarray}
where $O^{\Theta}_{lk}=\langle l|\Theta \hat{O} \Theta^{\dagger} |k\rangle$, $n$ corresponds to the real space and orbital coordinates and $l$ or $k$ corresponds to an arbitrary basis. For an arbitrary operator $\hat{O}$, we can find
\begin{eqnarray}\label{eqS7}
	O^{\Theta}_{lk} &=& \langle l|\Theta \hat{O} \Theta^{\dagger} |k\rangle
	=\sum_{i,j}\{\langle l|\Theta O_{ij} |i\rangle \langle j| \Theta^{\dagger} |k\rangle\}\nonumber\\
	&=&\sum_{i,j}\{O_{ij}^*\langle \Theta\Theta l| (-1) |\Theta i\rangle \langle \Theta j| (-1) |\Theta\Theta k\rangle\}\nonumber\\
	&=&\sum_{i,j}\{O_{ij}^*\langle i | \Theta l\rangle \langle \Theta k|j\rangle\}
	=\sum_{i,j}\{ \langle \Theta k|O_{ij}^*| j\rangle \langle i |\Theta l\rangle\}\nonumber\\
	&=& \langle \Theta k|O^{\dagger}|\Theta l\rangle=O^{\dagger}_{\Theta k,\Theta l}
\end{eqnarray}	
Substituting it into Eq.~\ref{eqS6},  we have
\begin{eqnarray}\label{eqS8}
	C_z (E_f,-V_E)&=& -2\pi i \sum_{n \in z\& l, k\in all }\{P_{\Theta l,\Theta n}^{\dagger}(V_E ) ((K^{(-x)\dagger}(V_E))_{\Theta k,\Theta l} (K^{(-y)\dagger}(V_E))_{\Theta n,\Theta k}\nonumber\\
	& &-(K^{(-y)\dagger}(V_E))_{\Theta k,\Theta l} (K^{(-x)\dagger}(V_E))_{\Theta n,\Theta k})\}\nonumber\\
	&=&-2\pi i \sum_{n \in z} \{\langle \Theta n|[K^{(-y)}_{E_f}(V_E),K^{(-x)}_{E_f}(V_E)]P_{E_f}(V_E)| \Theta n\rangle\}\nonumber\\
	&=&2\pi i \cdot \mathrm{Tr}\{ P_{E_f}(V_E ) [-i\lfloor -\hat{y},P_{E_f}(V_E)\rfloor ,-i\lfloor -\hat{x},P_{E_f}(V_E)\rfloor ]\}_{\mathcal{P}z}\nonumber\\
	&=& \sigma_{-y,-x}(z,E_f,V_E)/(e^2/h)=-\sigma_{xy}(z,E_f,V_E)/(e^2/h)\nonumber\\
	&=&-C_{\mathcal{P}z}(E_f,V_E)
\end{eqnarray}
Since $\Omega _z(\varepsilon)=d(C_z(\varepsilon))/d\varepsilon$,  we can conclude Eq.~\ref{eqS1}  holds.
Therefore, due to the  $\mathcal{P}\mathcal{T}$ symmetry, reversing electric fields can trigger  Berry- curvature  monopoles with opposite value on the inverse layer.

Furthermore, if we take disorder into consideration, we will have  $C_z(E_f,-V_E)[\omega]=C_z(E_f,-V_E)[\Theta\omega]$,
where $\omega=\omega ({\bf r},l)$ corresponds to one disorder configuration with a time-space $\mathcal{P} \mathcal{T}$ partner $\Theta\omega$.
Here ${\bf r}, l$ are the spatial and orbital indices. The even-layered AFM $\mathrm{MnBi_{2}Te_{4}}$ with the disorder
will  recover the $\mathcal{P} \mathcal{T} $ symmetry after averaging over disorder configurations $\omega$, because
the probability of $\omega$ and $\Theta\omega$ equal with $\rho(\omega)=\rho(\Theta\omega)$. So, Eq.~\ref{eqS1} also exists after averaging over disorder configurations $\omega$.
\section{Local current }
In this section, we illustrate the details of how we evaluate the local current. We use a two-terminal device
and apply a small voltage bias, $V_L-V_R$ between the left and right terminal of the device
to simulate the local current \cite{JHTAI,LHL3DQAH}.
By calculating the time derivative of the  electron number
and after some derivations, we can gain the currents between different sites
\begin{eqnarray}\label{eqS9}
		J_{\mathbf{i}} &=&e\left\langle\dot{N}_{\mathbf{i}}\right\rangle \nonumber\\
		&=&\frac{i e}{\hbar}\left\langle\left[H, \sum_{\alpha} N_{\mathbf{i}}\right]\right\rangle\nonumber\\
	    &=& \frac{2 e}{h} \sum_{\alpha, \beta,\mathrm{j}} \int_{-\infty}^{e V_{R}} d E \operatorname{Im}\left\{H_{\mathrm{i} \alpha, \mathrm{j} \beta}\left[G^{r}\left(\Gamma_{L}+\Gamma_{R}\right) G^{a}\right]_{\mathrm{j} \beta, \mathrm{i} \alpha}\right\}\nonumber \\
	    & &+\frac{2 e^{2}}{h} \sum_{\alpha,\beta,\mathrm{j}} \operatorname{Im}\left[H_{\mathrm{i} \alpha, \mathrm{j} \beta} G_{\mathrm{j} \beta, \mathrm{i} \alpha}^{n}\left(E_{F}\right)\right]\left(V_{L}-V_{R}\right)\nonumber \\
		&=&	\sum_{\alpha,\beta,\mathrm{j}} J_{\mathbf{i}\to \mathbf{j}},
\end{eqnarray}
where $J_{\mathbf{i}\to \mathbf{j}}$ is the current from the site $\mathbf {i}$ to the site $\mathbf {j}$.
$\Gamma_{L / R}=i\left[\Sigma_{L / R}^{r}-\Sigma_{L / R}^{a}\right]$
and $\Sigma^{r}_{L/R}(\Sigma^{a}_{L/R})$ is the retarded (advanced) self energy.
Besides $G^r(G^a)$ is the retarded (advanced) Green's function and
$G^n_{\mathbf{j}\beta,\mathbf{i}\alpha}=(G^r\Gamma G^a)_{\mathbf{j}\beta,\mathbf{i}\alpha}$.
Then the local current  between $n_s\times n_s \times 1$ supercells  $\tilde{J}_{\mathbf{i}\to \mathbf{j}}$ can be expressed as
\begin{eqnarray}\label{eqS10}
	\tilde{J}_{\mathbf{i}\to \mathbf{j}} &=& \sum_{\mathbf{m}\in \Omega_{\mathbf{i}}}\sum_{\mathbf{n}\in \Omega_{\mathbf{j}}}J_{\mathbf{m}\to\mathbf{n}},
\end{eqnarray}
where $\Omega_{\mathbf{i}}$ ($\Omega_{\mathbf{j}}$) denotes the supercell with the coordinate $\mathbf {i}$($\mathbf {j}$).
If we only take the nearest-neighbor hopping into consideration,
the current between the supercells can be expressed as
\begin{eqnarray}\label{eqS11}
	\tilde{J}_{\mathbf{i}\to \mathbf{i}+\mathbf{e}_l} &=& \sum_{n_1=0}^{n_s-1}\sum_{n_2=0}^{n_s-1} J_{[{\mathbf{r}_{\mathbf{i}}-\sum_{\mu} n_{\mu}(1-\delta_{\mu,l})\mathbf{e}_{\mu}}]\to [{\mathbf{r}_\mathbf{i}+\mathbf{e}_l-\sum_{\mu} n_{\mu}(1-\delta_{\mu,l})\mathbf{e}_{\mu}}]},
\end{eqnarray}
where $\mathbf{r}_{\mathbf{i}}=[i_1\cdot n_s, i_2\cdot n_s, i_3]$.
Next, let us evaluate the local current with varying the Fermi energy $E_f$ in the FM $\mathrm{MnBi_{2}Te_{4}}$.
\begin{figure}[htbp]
    \includegraphics[width=6.5in]{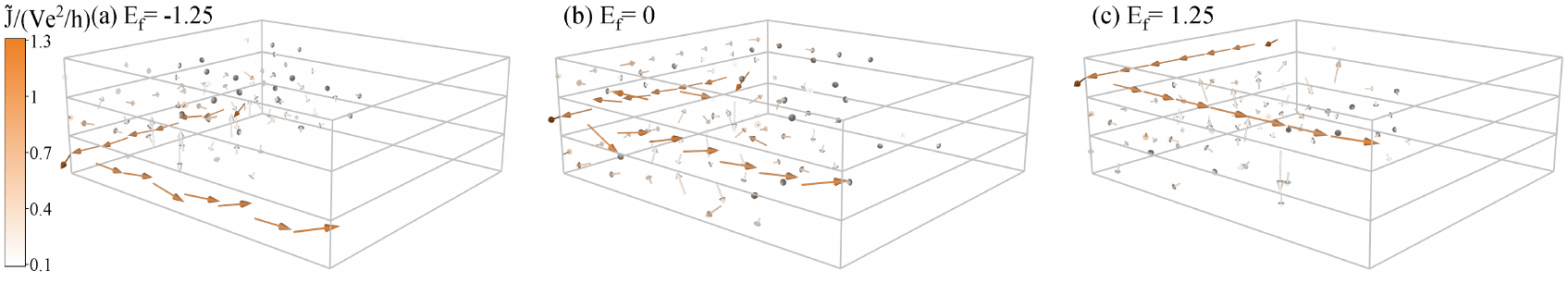}
    \caption{(color online) (a) Local current of the QALH in the FM phase when the Fermi energy varying from (a) $ E_f= -1.25$ to (b) 0 to (c) 1.25,
	with the E-field fixed $V_E = 1.2$ and system size $64 \times 64 \times 3$.
    }
    \label{figS3}
  \end{figure}
In Fig.~\ref{figS3} with $E_f=-1.25$, there exists only one chiral edge mode in the 1st layer
and the edge mode can be switched from the 1st layer to the 2nd layer and then to the 3rd layer
by varying the energy $E_f=-1.25$ to $0$ and then to $1.25$,
which is consistent with the schematic plots in Fig.~4(a) and numerical results in Fig.~4(c) in the main text. The local current in the bulk originates from the Anderson localized states induced by disorder.

\section{The experimental characterization of the layer-locked QAHE edge mode}
\begin{figure}[htbp]
    \includegraphics[width=6.5in]{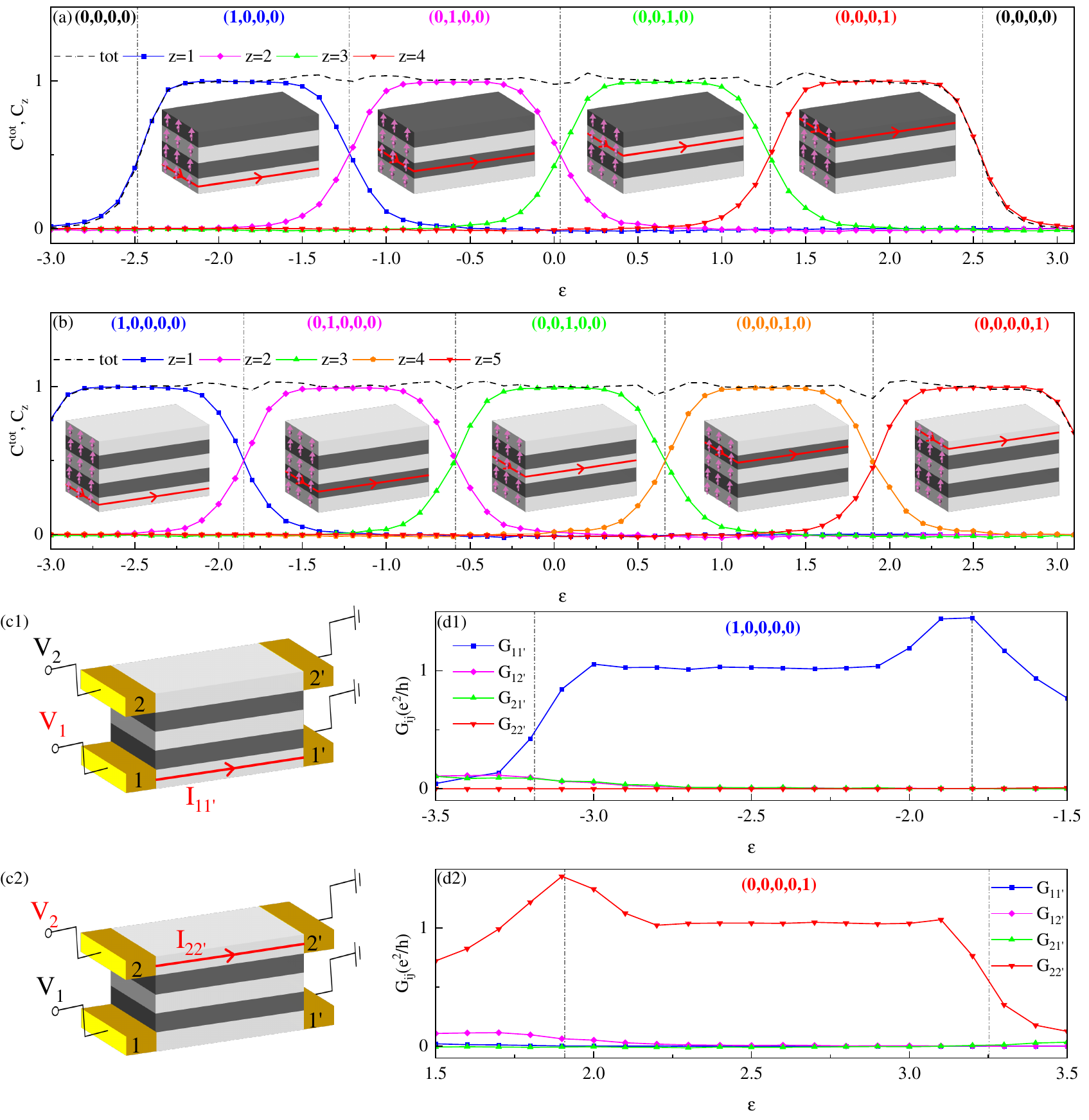}
    \caption{(color online) (a)-(b) $C^{tot}, C_z$ vs the Fermi energy $E_f$
	with E field fixed $V_E=1.25$ in the four-layered or five-layered FM $\mathrm{MnBi_{2}Te_{4}}$.
	(c1)-(c2) plot the schematic of the four-terminal device to detect the QALHE
	(d1)-(d2) plot $G_{ij}$ when the bottom/top layer is polarized with the QAHE edge mode
	and correspond to (c1)-(c2).
    }
    \label{figS4}
  \end{figure}
In this section, we will first show that the QALHE can also be realized in a much thicker system
and we take the phase transitions $(1,0,0,\cdots)\to (0,1,0,\cdots)\to (0,0,1,\cdots) $ in the FM $\mathrm{MnBi_{2}Te_{4}}$ as an example.
In Fig.~\ref{figS4}(a-b), we plot the Chern numbers as a function of energy $\varepsilon$, and find that such  phase transitions exist in four-layered and five-layered FM $\mathrm{MnBi_{2}Te_{4}}$ by varying the energy $\epsilon$, with the fixed electric field strength $V_E=1.25$.
Furthermore, we propose that the layer-locked QAHE edge mode can be detected by a four-terminal device as illustrated in Fig.~\ref{figS4}(c1-c2). Here, leads $1$ and $1'$($2$ and $2'$) are contacted to the left and right of the bottom (top) layer in order. According to the Landau-Buttiker formula,
the transmission coefficient between leads $i$ and $j$ can be expressed as
$T_{ij}=\mathrm{Tr}\{\Gamma_iG^r\Gamma_jG^a\} $, where $G^{r,a}$ is retarded (advanced) Green's function of the whole system and $\Gamma_{i}$ is the linewidth function of the $i$-th lead~\cite{datta_1995}, and the related differential conductance is $G_{ij}=e^2/hT_{ij}$.
When the five-layered FM is in the $(1,0,0,0,0)$ phase,
a QAHE chiral edge mode exists only in the bottom layer .
In this case,  because the bottom layer is turned on and other layers is turned off, 
a current flows from lead $1$ to lead $1'$ and current flow between other pair of leads, when the gate voltage $V_1$($V_2$) is applied on lead $1$(lead $2$) with other leads grounded.
As a result, one can see $G_{11'}=e^2/h$ and the other conductances are zero([see Fig.\ref{figS4}(d1)]).
On the other hand, in the $(0,0,0,0,1)$ phase, $G_{22'}=e^2/h$ and the other conductances are zero
since the chiral edge mode only exists in the top layer [see Fig.~\ref{figS4}(d2)].
%